%% file: main_letter.tex
\documentclass[aps, prl, 10pt, twocolumn, superscriptaddress, noshowpacs, preprintnumbers, floatfix]{revtex4-1}

\usepackage[colorlinks=true,breaklinks=true]{hyperref}
\hypersetup{allcolors=[rgb]{0.0 0.0 1.0},linkcolor=[rgb]{0.75 0.05 0.05}}
\usepackage[dvipsnames]{xcolor}
\usepackage{mathtools}
\usepackage{amsmath}
\usepackage{amsfonts}
\usepackage{amssymb}
\usepackage{bbold}
\usepackage{upgreek}
\usepackage{multirow}
\usepackage{bm}
\long\def\exclude#1{}
\usepackage{orcidlink}
\usepackage{slashed}

\DeclareMathOperator{\cm}{cm}

\DeclareMathOperator{\keV}{keV}

\DeclareMathOperator{\s}{s}

\DeclareMathOperator{\erg}{erg}

\DeclareMathOperator{\g}{g}

\newcommand{\bP}{{\bf P}}

\newcommand{\bp}{{\bf p}}

\newcommand{\bk}{{\bf k}}
\newcommand{\bq}{{\bf q}}

\newcommand{\bl}{{\bf l}}
\newcommand{\tk}{{\tilde{k}}}
\newcommand{\tl}{{\tilde{l}}}
\newcommand{\tP}{{\tilde{P}}}
\newcommand{\tp}{{\tilde{p}}}
\newcommand{\btp}{{\tilde{\mathbf{p}}}}

\newcommand{\te}{{\tilde{\varepsilon}}}
\newcommand{\tom}{{\tilde{\omega}}}
\newcommand{\tm}{{\tilde{m}_S}}
\newcommand{\bn}{{\bf n}}

\begin{document}

\title{Leading bounds on micro- to picometer fifth forces from neutron star cooling}

\author{Damiano F.\ G.\ Fiorillo \orcidlink{0000-0003-4927-9850}}
\affiliation{Deutsches Elektronen-Synchrotron DESY,
Platanenallee 6, 15738 Zeuthen, Germany}

\author{Alessandro~Lella~\orcidlink{0000-0002-3266-3154}}
\affiliation{Dipartimento Interateneo di Fisica  ``Michelangelo Merlin'', Via Amendola 173, 70126 Bari, Italy}
\affiliation{Istituto Nazionale di Fisica Nucleare - Sezione di Bari, Via Orabona 4, 70126 Bari, Italy}%

\author{Ciaran~A.~J.~O’Hare~\orcidlink{0000-0003-3803-9384}}
\affiliation{School of Physics, The University of Sydney, NSW 2006, Australia}

\author{Edoardo~Vitagliano~\orcidlink{0000-0001-7847-1281}}
\affiliation{Dipartimento di Fisica e Astronomia, Università degli Studi di Padova, Via Marzolo 8, 35131 Padova, Italy}
\affiliation{Istituto Nazionale di Fisica Nucleare (INFN), Sezione di Padova, Via Marzolo 8, 35131 Padova, Italy}

\begin{abstract}
The equivalence principle and the inverse-square law of gravity could be violated at short distances ($10^{-6}$ to $10^{-12}$ meters) by scalars sporting a coupling $g_N$ to nucleons and mass $\mathrm{eV}\lesssim m_\phi\lesssim\rm MeV$. 
We show for the first time that stringent bounds on the existence of these scalars can be derived from the observed cooling of nearby isolated neutron stars (NSs). Although NSs can only be used to set limits comparable to the classic supernova (SN)~1987A cooling bound in the case of pseudoscalars such as the QCD axion, the shallow temperature dependence of the scalar emissivity results in a huge enhancement in the effect of $\phi$ on the cooling of cold NSs.
As we do not find evidence of exotic energy losses, we can exclude couplings down to $g_N\lesssim 5 \times 10^{-14}$. Our new bound supersedes all existing limits on scalars across six orders of magnitude in $m_\phi$.
These conclusions also extend to Higgs-portal models, for which the bound on the scalar-Higgs mixing angle is $\sin\theta\lesssim 6\times 10^{-11}$. 

\end{abstract}


\maketitle

{\bf\textit{Introduction.}}---The existence of novel light CP-even scalars $\phi$ would lead to scale-dependent departures from firmly established predictions of gravitational physics, like Newton's inverse-square law and the weak equivalence principle. An experimental effort has been devoted to searching for deviations from these principles for several decades, see, e.g.,~\cite{Adelberger:2003zx,Will:2014kxa,Tino:2020nla}. Such scalar particles---which can arise as dilatons~\cite{Damour:1994zq,Taylor:1988nw} or radions in theories with extra dimensions~\cite{Arkani-Hamed:1999lsd}---could also constitute 100\% of the dark matter content in our Universe if produced through, e.g., the misalignment mechanism~\cite{Hui:2016ltb,Arvanitaki:2017nhi,Antypas:2022asj,Cyncynates:2024ufu,Cyncynates:2024bxw}; act as a portal with a rich dark sector~\cite{Knapen:2017xzo}; or be related to the hierarchy problem in the context of relaxion models~\cite{Flacke:2016szy}.

Casimir measurements~\cite{Sushkov:2011md,Chen:2014oda}, microcantilevers~\cite{Geraci:2008hb}, torsion-balance experiments~\cite{Kapner:2006si,Lee:2020zjt,Smith:1999cr,Smith:1999cr,Yang:2012zzb,Tan:2020vpf,Hoskins:1985tn}, and satellite-borne accelerometers~\cite{Berge:2017ovy} are powerful probes of scalars with masses $m_\phi\lesssim 1\,\rm eV$, corresponding to deviations from the inverse-square law at distances larger than $1\rm\, \upmu m$. Nonetheless, astrophysical bounds are much more stringent at larger masses. The existence of a new scalar particle is expected to affect the lives of stars. After being emitted by the nucleons making up the stellar plasma, the subsequent escape of the scalars out of the star constitutes an exotic cooling channel beyond what is usually assumed in standard evolutionary models~\cite{Grifols:1986fc,Grifols:1988fv,Raffelt:1996wa}.
Arguments of this type have been used in the past to set strong bounds on the scalar-nucleon coupling $g_N$. For example, the production of $\phi$ through resonant conversion from longitudinal plasmons would change the brightness of the tip of the red giant branch from what we observe, implying a severe limit of $g_N\lesssim 1.1\times 10^{-12}$~\cite{Hardy:2016kme}. Likewise, the exotic cooling due to scalar bremsstrahlung through electron-nucleus scattering would induce a deviation in the standard white-dwarf luminosity function, so that $g_N\lesssim 7\times 10^{-13}$~\cite{Bottaro:2023gep}. For $m_\phi\gtrsim 10\,\rm keV$ horizontal branch stars exclude a sliver of the parameter space~\cite{Hardy:2016kme}, and at larger masses still ($m_\phi\gtrsim 100 \,\rm keV$) the existence of scalars is constrained by the observed duration of the neutrino signal from supernova (SN)~1987A at Kamiokande II and IMB~\cite{Ishizuka:1989ts,Krnjaic:2015mbs,Hardy:2024gwy} (see however Ref.~\cite{Fiorillo:2023frv} for a recent comparison of standard neutrino cooling and SN~1987A data).

\begin{figure}
    \includegraphics[width=\linewidth]{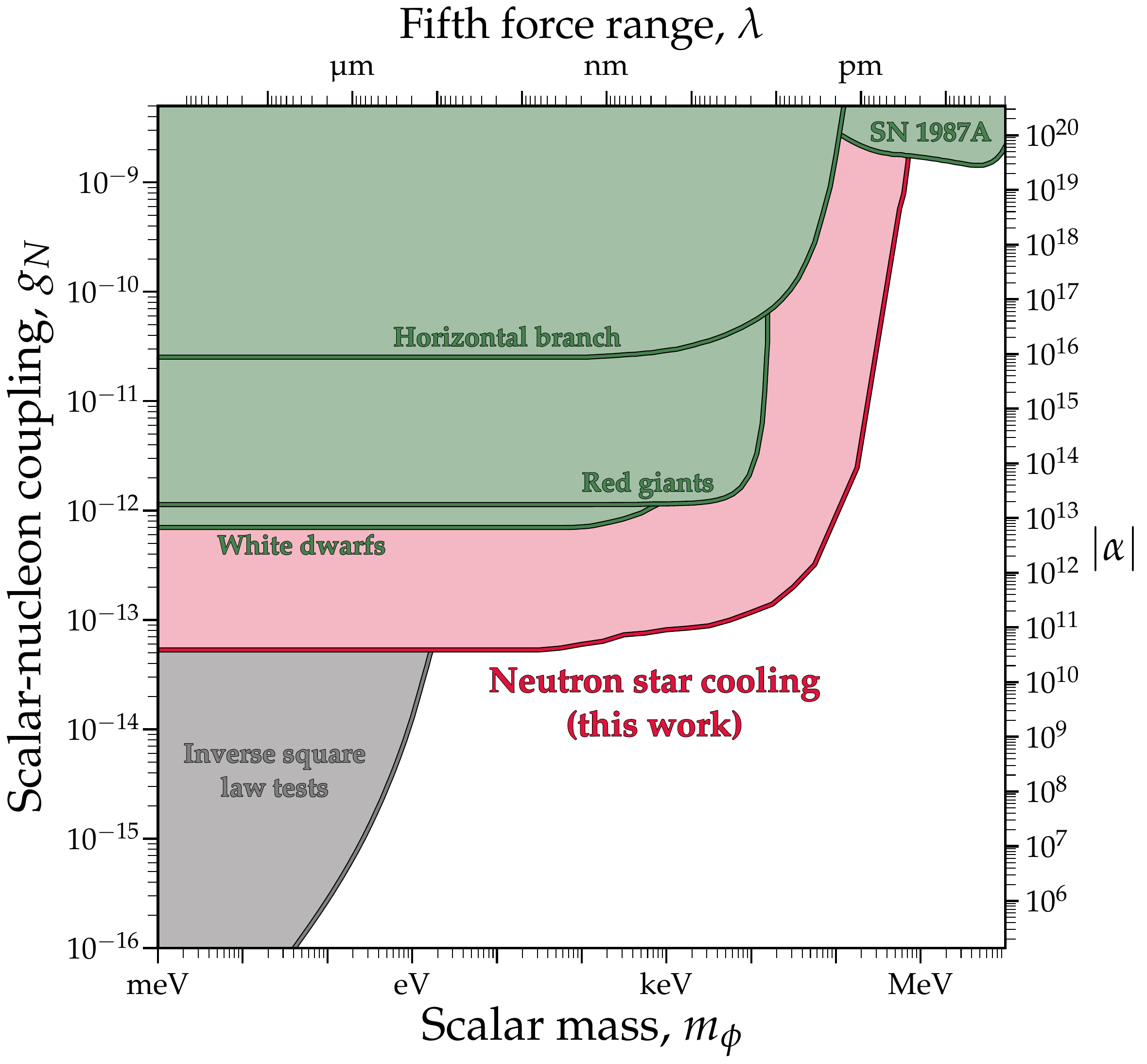}
    \caption{Our new limit on the scalar-nucleon coupling $g_N$ at 95\% CL as a function of the scalar mass from neutron star cooling. We show the limit alongside other astrophysical bounds (green)~\cite{Hardy:2016kme,Bottaro:2023gep,Hardy:2024gwy} and laboratory tests of the inverse-square law (gray)~\cite{Chen:2014oda}. For comparison with the conventional parametrization of fifth forces, we also show the coupling in terms of the equivalent strength of a Yukawa force relative to gravity, $|\alpha|$, along the right-hand axis, and the range of that force $\lambda = m_\phi^{-1}$ along the top axis. We assume equal coupling to protons and neutrons.}\label{fig1}
\end{figure}

For the first time, we show that old neutron stars (NSs), such as some of the nearby isolated Magnificent Seven~(M7) and PSR J0659, constrain $g_N\lesssim 5\times 10^{-14}$, improving existing bounds by more than one order of magnitude. Analogously to NS bounds placed on axions~\cite{Iwamoto:1984ir,Page:2010aw,Leinson:2014ioa,Sedrakian:2015krq,Hamaguchi:2018oqw,Buschmann:2021juv,Gomez-Banon:2024oux}, we derive bounds on scalar emission by comparing the predicted cooling curves of each NS with their measured ages and surface luminosities, fixing the equation of state (EoS) and the superfluidity model, while varying the uncertain NS mass and fraction of light elements in their envelopes. In contrast to the QCD axion, however, these constraints are dramatically stronger than those determined from SNe, as shown in Fig.~\ref{fig1}. At $m_\phi\simeq 100\,\rm keV$, our new bound is more than three orders of magnitude stronger than the one obtained from SN~1987A. Although this might come as a surprise, it can be understood through simple arguments that we outline below. The main cause is the lower temperature of cold NSs, which strongly suppresses the rate of neutrino emission in comparison to the rate of scalar emission, making it much easier to spot this signal of exotic cooling.

{\bf\textit{When are supernovae better than cold neutron stars?}}---The celebrated exotic cooling bounds from SNe, first drawn on the QCD axion, are the paradigmatic application of this kind of argument. Broadly speaking, the argument relies on the idea that we expect a dynamical impact on a star's evolution when the luminosity in the form of a new species becomes comparable to the dominant luminosity of the star. In the case of SNe this is due to neutrino emission, and for NSs---at least up to $\sim10^5\,$ years in age---this is still the case. This is why the use of cold NSs does not allow one to constrain any significant new portions of the QCD axion parameter space in comparison to SNe. In both cold NSs and hot proto-neutron stars formed in the core of a SN, the emission of axions and of neutrinos depends on the temperature roughly in the same way. This is because the emission of an axion or a neutrino pair comes with an associated flip of the spin of a nucleon. The amplitude for such a spin flip is always associated with a factor $\sim p_a/m_N$, as appears most naturally in the non-relativistic Hamiltonian for a spin flip, which has the form $\boldsymbol{\sigma}\cdot \bp_a/m_N$, where $\boldsymbol{\sigma}$ is the spin operator. Since the momentum of the axion, or of the neutrino pair, grows in proportion to the temperature $T$, it follows that the neutrino and axion luminosities scale in the same way with temperature, and thus their ratio, $L_a/L_\nu$, is roughly the same for SNe and cold NSs,
\begin{equation}
    \left(\frac{L_a}{L_\nu}\right)_{\rm NS}\sim \left(\frac{L_a}{L_\nu}\right)_{\rm SN} \, .
\end{equation}
This is why the values of the axion coupling constants that can be constrained are roughly the same for both of these sources.

Therefore, at present, cold NSs have primarily acted as an independent means, with complementary systematic uncertainties, to exclude coupling strengths comparable to the ones tested by SNe. The argument above provides a reason for this, but it also sheds light on its unique dependence on the axial mechanism of axion emission. Different emission mechanisms can exhibit characteristically different temperature dependencies, which we can understand with simple arguments. We focus on the case of a scalar particle coupling to nucleons here, which is the most promising target for neutron-star-like environments given the densities close to nuclear saturation. When emitting a scalar particle, the spin of the nucleon does \textit{not} flip, so there is no suppression from the scalar momentum. On the other hand, the emission of scalar radiation from a system of non-relativistic particles can happen only in proportion to their quadrupole moment, rather than to their charge. In the reaction $NN\to NN$ (we take identical nucleons for the moment), both the total charge (i.e.~number of nucleons) and current (i.e.~flux of nucleons) are conserved, with only their quadrupole moment allowed to change, which is required for the radiation of scalar particles. It follows that the radiation rate will be suppressed by a factor $\sim (p_N/m_N)^4$; crucially, this suppression factor depends on the nucleons' momentum, rather than the scalar's. In a SN, where nucleons are usually only mildly degenerate, their typical momentum is of order $p_N\sim \sqrt{m_N T}$, so coincidentally the suppression factor $\sim (p_N/m_N)^4\sim (T/m_N)^2$ turns out to be comparable to that for axion and neutrino emission. But in cold NSs, thanks to their large degeneracy, scalar emission is enhanced substantially in comparison to neutrinos. This simple argument suggests that,
\begin{equation}
    \left(\frac{L_\phi}{L_\nu}\right)_{\rm NS}\sim \left(\frac{L_\phi}{L_\nu}\right)_{\rm SN}\frac{p_{N,\rm NS}^4}{m_N^2 T_{\rm NS}^2}\sim 10^7 \left(\frac{L_\phi}{L_\nu}\right)_{\rm SN},
\end{equation}
where $T_{\rm NS}\sim 10\;\mathrm{keV}$ is the typical temperature for NSs with ages $\sim10^5\,$years and $p_{N,\rm NS}\sim 200\;\mathrm{MeV}$ is the typical nucleon Fermi momentum in nuclear matter at nuclear saturation density $\rho\sim\text{few}\times10^{14}\,\g\,\cm^{-3}$. Such a huge enhancement implies that NSs should be much better testbeds for the emission of scalar radiation than SNe. Since the luminosity grows in proportion to the squared coupling of the new species, we expect an improvement of around 3--4 orders of magnitude in the bounds on scalar particles from cold NSs.

To complement our arguments, we may also notice that if the new scalar particle couples with different strengths to protons and neutrons, one may well have an additional enhancement. While the total charge associated with the scalar field is conserved in $np\to np$ scattering, the current (i.e.~flux of protons or flux of neutrons) is not necessarily conserved. Therefore, one can have an even stronger enhancement, because the scalar radiation rate is suppressed only by a dipole factor $\sim (p_N/m_N)^2$, so that we may have, 
\begin{equation}
    \left(\frac{L_\phi}{L_\nu}\right)_{\rm NS}\sim \left(\frac{L_\phi}{L_\nu}\right)_{\rm SN}\frac{p_{N,\rm NS}^2}{ T_{\rm NS}^2}\sim 10^{8} \left(\frac{L_\phi}{L_\nu}\right)_{\rm SN}.
\end{equation}
These simple arguments, based on the microphysics of the emission process, point towards NSs as ideal probes of scalar emission. We now proceed to a discussion of how our resulting constraint, shown in Fig.~\ref{fig1}, is derived.

{\bf\textit{Scalars from cold neutron stars: data analysis and constraints}}---Under typical conditions expected for a NS with age $\sim10^5\,$years, the emission of scalars with masses $m_\phi\lesssim30\,\keV$ opens an efficient energy-loss channel in addition to neutrino and photon emission, which would accelerate the NS cooling rate. The total energy balance requires that,
\begin{equation}
    C \frac{dT_c^{\infty}}{dt}=-L_\gamma^\infty-L_\nu^\infty-L_\phi^\infty+H\, ,
\end{equation}
where the photon luminosity is related to the surface temperature $T_s^\infty$ as $L_\gamma^\infty=4\pi R_{*,\infty}^2\,(T_s^\infty)^4$. Here $t$ is the time, $C$ and $T_c^{\infty}$ are the heat capacity and temperature of the NS interior, respectively, and the superscript $\infty$ refers to quantities as measured by a distant observer. The final term $H$ accounts for surface heating due to the surrounding magnetic field. We assume here $H=0$--Ohmic heating from magnetic field decay might affect our constraints by $\sim50\%$, as shown in Ref.~\cite{Buschmann:2021juv}. Furthermore, we do not take into account other potential sources of stellar heating, e.g. rotochemical heating, vortex creep heating and crust cracking~\cite{Gonzalez:2010ta}, which might possibly induce uncertainties comparable to those related to Ohmic heating. It should be noted that the theoretical modeling of such effects still requires work and none of them has yet been firmly confirmed experimentally.

In the nucleon-rich interior of the NS, scalar production is dominated by the nucleon coupling $\mathcal{L}\supset g_N\phi \overline{N}N$, where $N=p,n$ for neutrons and protons, respectively. The main emission channel is $NN$ bremsstrahlung $N+N\rightarrow N+N+\phi$~\cite{Ishizuka:1989ts}. We re-evaluate the scalar bremsstrahlung emission rate here, going beyond the usual one-pion exchange~(OPE) or soft approximations employed in previous works. Nucleon scattering is described by employing the effective interaction potential already employed in Refs.~\cite{Friman:1979ecl,Bottaro:2024ugp} to compute neutrino emissivities from NSs.  The code we use to model the NS evolution, \texttt{NSCool}~\cite{2016ascl.soft09009P}, adopts the same framework to determine neutrino emission, so that our treatment of the exotic production is on par with that of standard processes. The potential has a long-range OPE interaction, reduced at large momentum exchange with a phenomenological $\rho$ meson exchange, and a short-range component extracted from the Landau parameters of the nuclear matter. We detail the computation of the scalar emission in the Supplemental Material~(SM)~\cite{supplementalmaterial}. 

If the scalar emission is too efficient, it competes with photon and neutrino cooling channels, potentially making the NSs cooler than expected for their age. We consider measurements of four of the M7~(see Refs.~\cite{Dessert:2019dos,Buschmann:2021juv} for related works on M7 NSs), for which thermal luminosity and kinematic age data are available~\cite{Potekhin:2020ttj,Suzuki:2021ium}. This list is enriched by PSR~J0659, which is also older than $\sim10^5$ years and has available thermal luminosity measurements. All the relevant data for our analysis are reported in Tab.~\ref{Tab:NSData}.
    
\begin{table}
\centering
\caption{Parameters of the neutron stars considered herein. Values of the core temperature $T_c$, the core density $\rho_c$ and the NS radius $R$ are obtained from the best-fit models determined for each given NS.}
\begin{tabular}{l c c c c c}
\hline
Name & $L_\gamma^\infty$ & Age & $T_c$ & $\rho_c$ & $R$\\
 & [$10^{33}$ erg/s] & [Myr] & [keV] & [g cm$^{-3}$] & [km]\\
\hline
J1856\textsuperscript{a} & $0.065 \pm 0.015$ & $0.42 \pm 0.08$ & $1.5$ & $7.1\times10^{14}$ & $11.4$  \\

J1308\textsuperscript{b} & $0.32 \pm 0.06$ & $0.55 \pm 0.25$ & $9.4$ & $1.2\times10^{15}$ & $11.2$ \\

J0720\textsuperscript{c} & $0.22 \pm 0.11$ & $0.85 \pm 0.15$ & $5.1$ & $1.1\times10^{15}$ & $11.3$ \\

J1605\textsuperscript{d} & $0.4 \pm 0.1$ & $0.44 \pm 0.07$ & $8.1$ & $1.2\times10^{15}$ & $11.2$\\

J0659\textsuperscript{e} & $0.28 \pm 0.14$ & $0.35 \pm 0.044$ & $9.0$ & $7.5\times10^{14}$ & $11.5$\\

\hline
\end{tabular}
\begin{minipage}{\linewidth}
\raggedright
\footnotesize
\textsuperscript{a}\cite{Mignani:2012mm, Ho:2006uk, Sartore:2012fk}; 
\textsuperscript{b}\cite{Motch:2009nq, Hambaryan:2011bu}; 
\textsuperscript{c}\cite{Tetzlaff:2011kh, Hambaryan:2017wvm}; 
\textsuperscript{d}\cite{Tetzlaff:2012rz, Pires:2019qsk}; 
\textsuperscript{e}\cite{Suzuki:2021ium, Zharikov:2021llh}.
\end{minipage}

\label{Tab:NSData}
\end{table}

We simulate the NS cooling curves using the public code {\tt NSCool}~\cite{2016ascl.soft09009P}, which tracks the cooling process of a one-dimensional NS model from a few seconds after its birth to several million years. In particular, {\tt NSCool} solves the heat transport and energy balance equations in a full General Relativity~(GR) framework, determining the surface temperature $T_s(T_c)$ as a function of the temperature in the NS interior $T_c$. The NS model is determined by specifying the NS mass $M_{\rm NS}$, the amount of light elements $\Delta M$ defining the envelope composition, the EoS modelling the NS interior, and the choice of superfluidity model. In the following we assume the Akmal - Pandharipande - Ravenhall~(APR) EoS~\cite{Akmal:1998cf}, built in the {\tt NSCool} code, and the {\tt 0-0-0} superfluidity model, which assumes no superfluidity by setting excitation gaps to zero; we can rely in this case on the detailed study of Ref.~\cite{Buschmann:2021juv}, showing that the constraints on exotic emission may change by tens of percent between different model assumptions, well below the unavoidable uncertainties on the nuclear interaction model. The latter reasonably entails a factor 5--6 uncertainty on the emissivity, corresponding to a factor $2-3$ uncertainty on the final constraints. 

Conversely, the fraction of light elements in accreted envelopes $\Delta M$ may significantly affect the relation between the core and surface temperatures $T_s(T_c)$ during both the neutrino- and photon-dominated cooling phases, while the baryonic NS mass has a major role in determining the composition and the density of the core. To deal with these nuisance parameters, we vary $\Delta M$ over six log-spaced values from $\Delta M=10^{-20}\,M_\odot$ (no light elements in accreted envelopes) to $\Delta M=10^{-6}\,M_\odot$ (high-concentration of light elements in accreted envelopes). For the NS mass, we take six linearly-spaced values between $M_{\rm NS} \in [1.0\,M_{\odot},2.0\,M_{\odot}]$. For each pair of $(\Delta M,M_{\rm NS})$, we run a set of NS simulations while varying the scalar's mass and coupling. The set of simulated light curves describing the $i^{\rm th}$ NS $L(m_\phi,g_N,{\bm \theta^i})$ is therefore be parametrized in terms of the scalar's free parameters $\{m_\phi$,$g_N\}$ and our nuisance parameters ${\bm \theta}^i=\{M_{\rm NS}^i,\,\Delta M^i,\,t^i\}$, which characterize the NS model. For each NS, we then write down a likelihood function,
\begin{equation}
    \begin{split}
        \mathcal{L}_i\left({\bm d}_i|m_\phi,g_N,{\bm \theta}^i\right)=\,&\mathcal{N}\left[L(m_\phi,g_N,{\bm \theta^i})-L_0^i,\,\sigma_L^i\right]\\
        &\times\mathcal{N}\left[t^i-t_0^i,\,\sigma_t^i\right] \, ,
    \end{split}
\end{equation}
where $\mathcal{N}(x,\sigma)$ is a zero-mean Gaussian distribution function with standard deviation $\sigma$. The data for the $i^{\rm th}$ NS consists of ${\bm d}_i=\{L_0^i,\,\sigma_L^i,\,t_0^i,\,\sigma_t^i\}$, where $\sigma_L^i$ and $\sigma_t^i$ are the 1$\sigma$ measurement uncertainties on the NS luminosity $L_0^i$ and age $t_0^i$, respectively. From here, we can then construct a joint likelihood for the set of all NSs, $\mathcal{L}\left({\bm d}|\,m_\phi,\,g_N,\,{\bm \theta}\right)$, by taking the product of each individual NS's likelihood, where now ${\bm d}=\{\bm{d}_i\}_{i=1}^5$ and ${\bm \theta}=\{\bm{\theta}^i\}_{i=1}^5$ represent the full lists of NS data and their nuisance parameters. 

To set upper limits on the allowed value of $g_N$, we calculate the profile likelihood ratio test statistic for exclusion limits~\cite{Cowan:2010js} at a fixed value of $m_\phi$,
\begin{equation}
    q(g_N ;m_\phi)= 
    \begin{cases}
        -2 \ln{\frac{\mathcal{L}\left({\bm d}|m_\phi,g_N,\hat{{\bm \theta}}\right)}{\mathcal{L}\left({\bm d}|m_\phi,\hat{g}_N,\hat{{\bm \theta}}\right)}} & g_N\geq\hat{g}_N\,,\\
        0 & g_N<\hat{g}_N\,,
    \end{cases}
\end{equation}
where $\hat{g}_N$ is the best-fit value of $g_N$ and we use $\hat{\bm{\theta}}=\left\{\hat{\bm{\theta}}^i \right\}_{i=1}^5$ to denote the values of the nuisance parameters that maximize the likelihood function in which they appear.
The 95\% confidence level upper limit on the scalar-nucleon coupling can then be found by satisfying $q(\,g_N^{95};m_\phi)=2.71$, assuming that Wilks' theorem holds~\cite{Wilks:1938dza}. 

We note that the analysis of axion-induced cooling of NSs in Ref.~\cite{Buschmann:2021juv} found that a full Monte Carlo simulation of the distribution of $q$ differed from that of the asymptotic $\chi^2$ distribution, which is assumed when naively applying Wilks' theorem---this entailed a $\sim 50\%$ shift in the resulting upper limit on the axion mass. However, we emphasize that this is a minor discrepancy when compared to the dominant systematic uncertainties due to the nuclear interaction model, which are at the level of a factor 4--5 in the emissivity and therefore a factor 2 in the constraints. 

Our nuclear interaction model follows the historical setup of Ref.~\cite{Friman:1979ecl}, in which the long-range interaction is modeled as a OPE, potentially modified by a phenomenological $\rho$ meson exchange, and the short-range interaction is extracted from the Landau parameters of heavy nuclei. However, such Landau parameters refer to symmetric nuclear matter, whereas NSs are largely neutron dominated. Even the long-range component of the potential has been shown to suffer large renormalization in dense nuclear matter~\cite{Schwenk:2003pj,Schwenk:2003bc, Shternin:2018dcn, vanDalen:2003zw, Bacca:2008yr, DehghanNiri:2016cqm, Blaschke:1995va}. The quenching of the axion coupling in dense matter also cannot be accurately anticipated; the Brown-Rho scaling, adopted in e.g.~Ref.~\cite{Buschmann:2021juv}, is mostly a qualitative conjecture to capture the leading features of this phenomenon. Therefore, a fair assessment is that these techniques primarily capture the order of magnitude and the temperature dependence of the emissivity, and are the dominant source of uncertainty on the constraints we find, making efforts to correct for more minor 10--50\% discrepancies unnecessary. Fortunately, our constraints supersede previous ones by much more than these estimated uncertainties.

Our resulting limit is shown in Fig.~\ref{fig1}. We exclude $g_N\lesssim 5 \times 10^{-14}$ for masses $m_\phi\lesssim100\rm \, eV$, overtaking previous constraints from the WD luminosity function~\cite{Bottaro:2023gep} by more than one order of magnitude. Remarkably, our limits dominate over all the previous astrophysical limits for $m_\phi\lesssim 700\,\rm keV$, improving upon the SN cooling bound by $\sim 4$ orders of magnitude. For the SN constraints, we use the results of Ref.~\cite{Hardy:2024gwy}; these rely primarily on resonant emission, rather than bremsstrahlung emission; however, the two are comparable in order of magnitude, so that our estimates for the expected improvement in the NS cooling constraints still hold.

{\bf\textit{Discussion and outlook}}---At the microscopic level, searches for deviations from the inverse-square law and the weak equivalence principle can be seen as testing for novel Yukawa interactions arising from the exchange of low-mass bosons. By capitalizing on the low temperatures and high densities of cooling isolated NSs, we have obtained the most stringent bounds to date on exotic scalars $\phi$ with mass $\mathrm{eV}\lesssim m_\phi\lesssim \mathrm{MeV}$ coupled to nucleons. We exclude couplings down to $g_N\lesssim 5 \times 10^{-14}$, which in terms of the equivalent strength of the Yukawa force with respect to gravity means $|\alpha|=g_N^2/4\pi G_Nm_N^2\lesssim 4\times 10^{10}$, with an improvement over the previous bounds on the latter by more than two orders of magnitude. Our bounds are now leading for scalars mediating a fifth force over micro- to picometer distances---an especially challenging range to probe experimentally---emphasizing the complementary nature of astrophysical probes of new physics. 

These bounds are also relevant for CP-violating axion interactions~\cite{Moody:1984ba, Georgi:1986kr, Pospelov:1997uv,Ellis:1978hq,Khriplovich:1985jr,Gerard:2012ud,Okawa:2021fto, Pospelov:2005pr,Plakkot:2023pui,Dekens:2022gha,DiLuzio:2023cuk,Bigazzi:2019hav,Bertolini:2020hjc,DiLuzio:2023lmd,DiLuzio:2024ctr}, as we detail in the SM~\cite{supplementalmaterial}, where we compare our bound against the landscape of constraints from Refs.~\cite{Berge:2017ovy,Berge:2021yye,MICROSCOPE:2022doy,Smith:1999cr,Kapner:2006si,Lee:2020zjt,Ke:2021jtj,Tu:2007zz,Yang:2012zzb,Tan:2020vpf,Tan:2016vwu,Yang:2012zzb,Chen:2014oda,Capozzi:2020cbu,Heckel:2008hw,Crescini:2017uxs,Crescini:2016lwj,Wineland:1991zz,Fan:2023hci,Lee:2018vaq,Agrawal:2022wjm,Agrawal:2023lmw,Wu:2023ypz,Venema:1992zz,Lee:2018vaq,Tullney:2013wqa,Feng:2022tsu,Wei:2022ggs,Arvanitaki:2014dfa, Geraci:2017bmq,Blakemore:2021zna,Venugopalan:2024kgu,Chiu:2009fqu,Bezerra:2010pq,Sushkov:2011md,Banishev:2012kkb,Banishev:2014jka,Mostepanenko:2020lqe,Klimchitskaya:2013rwd,Klimchitskaya:2023cgy,Pokotilovski:2006up,Haddock:2017wav,Heacock:2021btd,Kamiya:2015eva,Bogorad:2023zmy,Raffelt:2012sp,OHare:2020wah,AxionLimits,Stadnik:2017hpa,Baruch:2024fbh,Baruch:2024frj,Carenza:2021pcm,Ferreira:2022xlw,Fiorillo:2025sln,Liu:2025ows, Grossman:2025cov}. We may also translate our limits to e.g.~models with a Higgs portal where $g_N\simeq 8\times 10^{-4}\sin\theta$. Our bound of $\sin\theta\lesssim 6\times 10^{-11}$ surpasses the existing limits that rely on the scalar-electron coupling. 

While in this work we have focused on scalars, fifth forces between baryons might be mediated by new gauge bosons---although in this case, one would need an additional coupling to leptons to guarantee anomaly cancellation. NSs might also constitute our best laboratory for exploring this class of models. Likewise, NSs can be the ideal probe of dark sector particles produced through baryonic interactions. We leave the question of how these cold compact stars fare compared to other probes to future work.

{\bf\textit{Acknowledgments.}}---We are grateful to Edward Hardy, Anton Sokolov, and Henry Stubbs for useful comments on an early version of this Letter. This article is based upon work from COST Action COSMIC WISPers (CA21106), 
supported by COST (European Cooperation in Science and Technology). DFGF is supported by the Ale\-xander von Humboldt Foundation (Germany). The work of AL is partially supported by the research grant number 2022E2J4RK ``PANTHEON: Perspectives in Astroparticle and
Neutrino THEory with Old and New messengers'' under the program PRIN 2022 (Mission 4, Component 1, CUP I53D23001110006) funded by the Italian Ministero dell'Universit\`a e della Ricerca (MUR) and by the European Union – Next Generation EU. 
CAJO is supported by the Australian Research Council under the grant numbers DE220100225 and CE200100008.
 EV is supported by the Italian MUR Departments of
Excellence grant 2023-2027 ``Quantum Frontier'', by the Italian MUR through the FIS 2 project FIS-2023-01577 (DD n. 23314 10-12-2024, CUP C53C24001460001), and by Istituto Nazionale di Fisica Nucleare (INFN) through the Theoretical Astroparticle Physics (TAsP) project.

\bibliographystyle{bibi}
\bibliography{references}
\onecolumngrid

\include{SMmod.tex}

\end{document}

%% file: SMmod.tex
\onecolumngrid
\appendix

\setcounter{equation}{0}
\setcounter{figure}{0}
\setcounter{table}{0}
\setcounter{page}{1}
\makeatletter
\renewcommand{\theequation}{S\arabic{equation}}
\renewcommand{\thefigure}{S\arabic{figure}}
\renewcommand{\thepage}{S\arabic{page}}

\begin{center}
\textbf{\large Supplemental Material for the Letter\\[0.5ex]
{\em Leading bounds on micro- to picometer fifth forces from neutron star cooling}}
\end{center}

\bigskip

In this Supplemental Material, we collect our results concerning production of scalars in cold neutron stars, additional information on neutron star observations, and an update of current bounds on CP-violating axions.

\bigskip

\twocolumngrid

\section{A.~Scalar emission from nuclear bremsstrahlung in cold neutron stars}

The emission of new particles from nuclear bremsstrahlung is affected by unavoidable uncertainties, especially in neutron-dominated matter. In the case of scalar particles, most historical treatments~\cite{Ishizuka:1989ts} have focused on a modeling of the nuclear interaction in terms of one-pion-exchange (OPE) alone, which however is not an accurate depiction of the short-range components of nuclear scattering even in vacuum. Here, we follow through the treatment introduced in Ref.~\cite{Friman:1979ecl}, which pivots around a Fermi liquid assumption for the nuclear matter. We introduce, however, the corrections recently adapted for the emission of neutrinos in Ref.~\cite{Bottaro:2024ugp}. Thus, the nuclear interaction amplitude is modeled as a long-range OPE contribution, which at large momentum exchange is phenomenologically suppressed by the introduction of a $\rho$-meson-exchange potential, tuned to reproduce the tensor channel nuclear effective potential as in Ref.~\cite{Ericson:1988wr}. In addition, we model the short-range interaction of nucleons in the medium by means of the Landau parameters, which represent the amplitudes for nucleon forward scattering; following Ref.~\cite{Friman:1979ecl}, these are extracted from the compressibility and response of the nuclear medium. 

The bremsstrahlung process we consider is $N_1+N_2\to N_3+N_4+\phi$, where $N_i$ is a nucleon and $\phi$ is the scalar field. Each nucleon $N_i$ has  four-momentum $p_i$,
momentum $\bp_i$, and an effective mass $m_N$, not necessarily equal between protons and neutrons. The proton-neutron mass difference is some tens percent, comparable with the level of approximation introduced by the many other assumptions (e.g. non-relativistic nucleons, recoilless emission), and therefore we neglect it. Nucleons have a kinetic energy $\varepsilon_i=|\bp_i|^2/2m_N$, whereas their rest energy is absorbed in the definition of the chemical potential. The momentum and energy of $\phi$ are denoted by $\bq$ and $\omega=\sqrt{|\bq|^2+m_a^2}$ respectively, with four-vector $q$. Our primary approximation is that $\bq\ll\bp_i$ and can therefore be neglected when compared with the nucleon momenta; on the other hand $\omega\sim \varepsilon_i$ are of the same order, comparable with the temperature, and therefore the scalar energy must be kept.

For convenience, in the matrix element we consider the nucleon spinors to be normalized by the non-relativistic condition $\overline{u}_i u_i=1$, rather than the more conventional relativistic condition $\overline{u}_i u_i=2 \varepsilon_i\simeq 2m_N$. In these conditions, we can directly write the emissivity for the scalar (number of particles emitted per unit time per unit volume per unit energy) as
\begin{widetext}
\begin{equation}\label{eq:master_integral}
    {\cal N}_s=\frac{dN_s}{dt dV d\omega}=\frac{\omega |\bq|}{2\pi^2}\prod_i \int \frac{d^3\bp_i}{(2\pi)^3}\frac{1}{2\omega}(2\pi)^4\delta^3\left(\sum_i \bp_i\right) \delta\left(\varepsilon_1+\varepsilon_2-\varepsilon_3-\varepsilon_4-\omega\right)S f_1 f_2 (1-f_3)(1-f_4)|\mathcal{M}|^2,
\end{equation}
\end{widetext}
where $f_i=f^{\rm FD}(\varepsilon_i-\mu_i)$ is the Fermi-Dirac function evaluated at the nucleon energy, and $\mu_i$ is the nucleon chemical potential. The symmetry factor $S$ depends on the identity of the nucleons involved in the process, i.e., $S=1/4$ for $nn$ and $pp$ scattering, and $S=1$ for $np$ scattering. This difference alone suggests that $np$ scattering typically dominates. The matrix element $|\mathcal{M}|^2$ is assumed to be summed over all spins, both for initial and final particles.

\subsection{Phase-space integration}\label{sec:phase_space}

First, we express the emissivity in terms of phase space integrals that can be simplified in the degenerate regime of neutron stars. We assume that the matrix element depends only on the momentum transfers between the nucleons $\bk=\bp_1-\bp_3=\bp_4-\bp_2$ and $\bl=\bp_1-\bp_4=\bp_3-\bp_2$. For reference, we will do all calculations for $np$ scatterings, assuming that $\bp_1$ and $\bp_3$ are the proton momenta, and $\bp_2$ and $\bp_4$ are the neutron momenta. For $nn$ or $pp$, the results can be simply obtained by taking the Fermi momenta to be identical.

The distribution function to be integrated over depends on the species; thus, for reference we will do all the calculations for $np$ scattering, where $\bp_1$ and $\bp_3$ are proton momenta, and $\bp_2$ and $\bp_4$ are neutron momenta. For $nn$ or $pp$ the results can be simply obtained by taking the chemical potential of the two species equal to a common value.

In Eq.~\eqref{eq:master_integral}, we apply the parameterization $\bp_1=\bP+(\bk+\bl)/2$, $\bp_2=\bP-(\bk+\bl)/2$, $\bp_3=\bP-(\bk-\bl)/2$, $\bp_4=\bP+(\bk-\bl)/2$, so that $\bP$ is the center-of-mass momentum, $\bk$ is the momentum exchanged in $t$-channel, and $\bl$ is the momentum exchanged in $u$-channel. In turn, all the integrals can be expressed
in terms of the three modules $P$, $k$, $l$, and three angles: $\theta$, the angle between $\bl$ and $\bk$; $\alpha$, the angle between $\bP$ and $\bk$; $\beta$, the azimuthal angle between $\bP$ and the plane containing $\bk$ and $\bl$. Only five variables are independent; the angle $\theta$ is obtained from the condition of energy conservation
\begin{equation}\label{eq:angle_kl}
    kl\cos\theta=m_N\omega.
\end{equation}
With this parameterization, after integrating the delta function, we obtain

\begin{widetext}
\begin{equation}\label{eq:parameterized_Ns}
    \mathcal{N}_s=\frac{|\bq| m_N}{128\pi^8}\int k dk l dl P^2 dP d\cos\alpha d\beta S f_1 f_2 (1-f_3)(1-f_4) |\mathcal{M}|^2;
\end{equation}
notice that the integral depends on $\omega$ after the condition on $\theta$ is enforced everywhere.
We now make all momenta dimensionless, denoting them by a tilde as $\tilde{k}=k/\sqrt{m_N T}$. Similarly, we use a dimensionless variable $\tilde{\omega}=\omega/T$, and in place of the chemical potentials we use the dimensionless potentials $\eta_i=\mu_i/T$, so that we find
\begin{equation}\label{eq:expression_in_integrals}
    \mathcal{N}_s=\frac{|\bq| m_N \left(m_N T\right)^{7/2}}{128\pi^8} \int \tk d\tk \tl d\tl \tP^2 d\tP d\cos\alpha d\beta S f_1 f_2 (1-f_3)(1-f_4) |\mathcal{M}|^2.
\end{equation}

We can now rewrite this expression in a more symmetric form as
\begin{equation}\label{eq:master_integral_symmetric}
    \mathcal{N}_a=\frac{|\bq| m_N\left(m_N T\right)^{7/2}}{128\pi^8}\frac{1}{8\pi^2}\int \prod_i d^3\tilde{\bp}_i \delta\left(\te_1+\te_2-\te_3-\te_4-\tom\right) \delta\left(\sum_i\tilde{\bp}_i\right) f_1 f_2 (1-f_3) (1-f_4) |\mathcal{M}|^2.
\end{equation}

Both the proton and the neutron distribution functions are rapidly varying close to their Fermi surface, so we can integrate all their momenta via the identity
\begin{equation}
    \int d\te_1 d\te_2 d\te_3 d\te_4 \delta(\te_1+\te_2-\te_3-\te_4-\tom) f_1 f_2 (1-f_3) (1-f_4)=J(\tom),
\end{equation}
with 
\begin{equation}
    J(\tom)=\frac{2\pi^2\tom}{3(e^\tom-1)}\left(1+\frac{\tom^2}{4\pi^2}\right).
\end{equation}

Thus we find
\begin{equation}\label{eq:integral_strongly_degenerate}
    \mathcal{N}_S=\frac{|\bq| m_N \left(m_N T\right)^{7/2}J(\tom)}{64\pi^{10}}\int \prod_i d^3\btp_i \delta(\tp_i^2-\tp_{F,i}^2) \delta\left(\sum\btp_i\right) |\mathcal{M}|^2,
\end{equation}
where $\tp_{F,1}=\tp_{F,3}=\tp_p=\sqrt{2\eta_p}$ is the proton Fermi momentum and $\tp_{F,2}=\tp_{F,4}=\tp_n=\sqrt{2\eta_n}$ is the neutron Fermi momentum.

The integral can now be simplified in terms of the original parameterization used for Eq.~\eqref{eq:parameterized_Ns}. The four delta functions enforce four kinematic conditions, which translate into
\begin{equation}\label{eq:kinematic_conditions}
    \theta=\frac{\pi}{2},\;\alpha=\frac{\pi}{2},\; \cos\beta=\frac{\tp_p^2-\tp_n^2}{2\tl \tP},\;\tP^2=\frac{\tp_p^2+\tp_n^2}{2}-\frac{\tk^2+\tl^2}{4}.
\end{equation}
Notice that these conditions admit two independent solutions for $\beta$ with opposite sign, so the overall integral must be multiplied by $2$ to account for both solutions. The final result reads
\begin{equation}\label{eq:N_a_final}
    \mathcal{N}_s=\frac{|\bq| m_N \left(m_N T\right)^{7/2} J(\tom)}{32\pi^8}
    \int \frac{d\tk d\tl}{\sqrt{2(\tp_n^2+\tp_p^2)-\tk^2-\tl^2-\frac{(\tp_n^2-\tp_p^2)^2}{\tl^2}}} |\mathcal{M}|^2,
\end{equation}

where the integration region is chosen so as to guarantee that the kinematic conditions in Eq.~\eqref{eq:kinematic_conditions} can be satisfied.

\end{widetext}

\begin{widetext}

\subsection{Matrix element for scalar emission}

We now determine the squared matrix element $|\mathcal{M}|^2$. This requires primarily to model the interaction potential between nucleons. Notice that the vacuum nuclear potential would here be inappropriate, as the properties of nucleons are strongly renormalized by the dense medium. In fact, in such a medium, nucleons should really be interpreted as quasi-particles in the Landau sense. Their long-range interaction, which is primarily mediated by the pion, the lightest meson, can be approximately taken to be the same as in vacuum. This should be regarded as an approximation up to factors of order unity, as the medium polarization can in reality renormalize the long-range interaction; see the discussion in Refs.~\cite{Schwenk:2003pj,Schwenk:2003bc, Shternin:2018dcn, vanDalen:2003zw, Bacca:2008yr, DehghanNiri:2016cqm, Blaschke:1995va}. The short-ranged interaction can instead be modeled by the Landau parameters, which define the forward scattering amplitude for nucleons; for a short-ranged potential, this single number is sufficient to define the interaction amplitude for any scattering angle. The Landau parameters are inferred from measurements of the response functions of heavy nuclei, though one should stress that such measurements are usually performed for symmetric nuclei, whereas nuclear matter in neutron stars is strongly asymmetric. Nevertheless, in order to stick with a definite framework, here we focus on the nuclear potential adopted in Ref.~\cite{Bottaro:2024ugp}. Overall, this nuclear potential is parameterized as
\begin{equation}\label{Eq:GenPotential}
V(\mathbf{k}) = f + f' \boldsymbol{\tau}_1 \cdot \boldsymbol{\tau}_2 + g \boldsymbol{\sigma}_1 \cdot \boldsymbol{\sigma}_2 + g'_\bk \boldsymbol{\tau}_1 \cdot \boldsymbol{\tau}_2 \, \boldsymbol{\sigma}_1 \cdot \boldsymbol{\sigma}_2  + h'_\bk  (\boldsymbol{\sigma}_1 \cdot \boldsymbol{\hat{\bk}}) (\boldsymbol{\sigma}_2 \cdot \boldsymbol{\hat{\bk}}) \boldsymbol{\tau}_1 \cdot \boldsymbol{\tau}_2.
\end{equation}

The spin-spin interaction $g'_\bk$ arises both from the long-range $\rho$ meson exchange and from the Landau parameters
\begin{equation}\label{eq:gk}
    g'_\bk \equiv g' - C_\rho \, \frac{f_\pi^2}{m_\pi^2} \frac{|\bk|^2}{|\bk|^2+m_\rho^2},
\end{equation}
where we take $m_\rho = 769 \, \rm MeV$, $C_\rho = 1.4$, and  $f_\pi \simeq 1$. 
For the Landau parameters instead we have $\{f, f', g, G\} = \frac{\pi^2}{2 \, m_{\rm N} \, p_{\rm F}(n)} \{F_0, F_0', G_0, G'_0\}$, where $F_0=0.5, F_0' = 0.7, G_0 = G_0' = 1.1$ and where $p_{\rm F}$ is the neutron Fermi momentum. Finally, the tensor interaction is
\begin{equation}\label{eq:hk}
    h'_\bk \equiv - \frac{f_\pi^2}{m_\pi^2} \Big(\frac{|\bk|^2}{|\bk|^2 +m_\pi^2} - C_\rho \frac{|\bk|^2}{|\bk|^2 +m_\rho^2}\Big).
\end{equation}

We highlight that the framework chosen for the computation sets scalar and neutrino emissivities on the same level of precision. In particular, we neglect possible effects due to scalar emission from the mediator of nucleon interaction~\cite{Dev:2020eam,Hardy:2024gwy}, which is expected to be subdominant in the nucleon degenerate medium of neutron stars. Indeed, we do not believe that any precision statement (i.e. more than an order-of-magnitude estimate) can be made in regards to the amount of emitted scalar particles.

Regarding the matrix element, we assume an interaction Lagrangian between the scalar particles and the nuclei of the form
\begin{equation}
    \mathcal{L}=-g_p \overline{p}p-g_n \overline{n}n=-g\overline{N}N-\delta g\overline{N}\tau^z N,
\end{equation}
where $N$ is the nucleon doublet and $\tau^z$ is the third component of the isospin matrix. We are also denoting $g=(g_p+g_n)/2$ and $\delta g=(g_n-g_p)/2$. Let us also introduce the one-particle operator $G=g+\delta g \tau^z$. 

The matrix element can now be determined from the bremsstrahlung diagram. We are going to obtain that by assuming nucleons to be non-relativistic; however, we will maintain the terms of order $v^2/c^2$, where $v$ is the nucleon velocity, since for $nn$ and $pp$ bremsstrahlung they are the dominant terms. This is as expected, since scalar emission from particles with equal charges appears primarily at the quadrupole level. For $np$ bremsstrahlung, if $\delta g\neq 0$, we can have scalar emission already at the monopole level; our calculation below confirms these insights.

\begin{figure*}
    \includegraphics[width=0.6\textwidth]{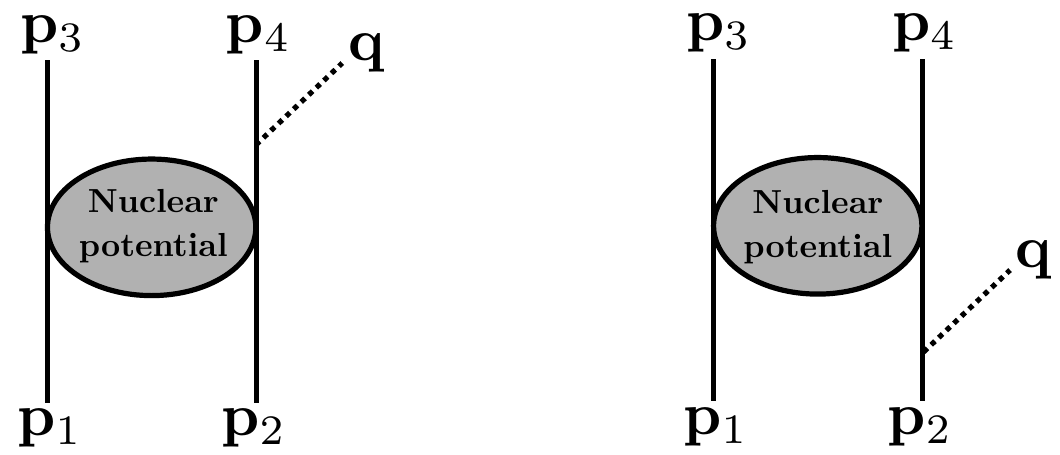}
    \caption{Schematic pair of Feynman diagrams contributing to nucleon-nucleon bremsstrahlung. The dark bubble represents an insertion of the nuclear interaction potential.}\label{fig:feyn_diag}
\end{figure*}

The diagrams contributing to nucleon-nucleon bremsstrahlung can be organized in pairs, each of which has the form of Fig.~\ref{fig:feyn_diag}; let us consider this pair first. The nuclear potential leads to an insertion of the form $V(\bk) \mathcal{O}_1 \mathcal{O}_2$, where $V_{\mathcal{O}}(\bk)$ is a component of the interaction potential depending on the momentum exchange $\bk$, and $\mathcal{O}_1$ and $\mathcal{O}_2$ are the corresponding operators acting on each of the two nucleons. For example, for the spin-spin component of the interaction potential, this will take the form $V_\sigma(\bk) \boldsymbol{\sigma}_1\cdot \boldsymbol{\sigma}_2$. Our notation in terms of generic operators $\mathcal{O}_1$ and $\mathcal{O}_2$ allows us to obtain the result for all components of the interaction potential at once in a schematic fashion. Notice that for both diagrams the potential interaction $V(\bk)$ depends on the momentum exchange $\bk=\bp_3-\bp_1\simeq\bp_2-\bp_4$, since $|\bq|\ll |\bk|$. The amplitude for the two diagrams in Fig.~\ref{fig:feyn_diag} is therefore
\begin{equation}
    \mathcal{M}^{(1)}=V_{\mathcal{O}}(\bk) \overline{u}_3 \mathcal{O} u_1\left[\frac{\overline{u}_4 G (\slashed{p}_4+\slashed{q}+m_N)\mathcal{O}u_2}{(p_4+q)^2-m_N^2}+\frac{\overline{u}_4 \mathcal{O} (\slashed{p}_2-\slashed{q}+m_N) G u_2}{(p_2-q)^2-m_N^2}\right].
\end{equation}

This expression can be expanded in powers of $v_F$; we can neglect everywhere terms of order $\omega/m_N\ll v_F$ since $\omega\sim T$ while $v_F\sim \sqrt{\mu/m_N}\sim 0.1$. This means that in the numerators we can neglect $\slashed{q}$ and set $\slashed{p}_4+m_N\simeq \slashed{p}_2+m_N\simeq 2m_N$ when acting on the on-shell non-relativistic spinors. In the denominators we can neglect the terms $q^2=m_a^2$ and expand the propagators to second order. The final result is, for $pp$ and $nn$ scattering, and introducing the unit direction $\bn=\bq/|\bq|$,
\begin{equation}\label{eq:quadrupole}
    \mathcal{M}=\frac{g_N}{m_N^2 \omega}\sum_\mathcal{O}\left(V_{\mathcal{O}}(\bk) \overline{u}_3\mathcal{O}u_1\overline{u}_4\mathcal{O}u_2-V_{\mathcal{O}}(\bl) \overline{u}_4\mathcal{O}u_1\overline{u}_3\mathcal{O}u_2\right) \Delta Q_{ij} n_i n_j,
\end{equation}
where
\begin{equation}
    \Delta Q_{ij}=p_{3,i}p_{3,j}+p_{4,i}p_{4,j}-p_{1,i}p_{1,j}-p_{2,i}p_{2,j}
\end{equation}
is the change in the (traceless) quadrupole tensor during the scattering; notice that $\Delta Q_{ii}=0$ to lowest order in $\omega/m_N$, due again to energy conservation. We can also rewrite this identically as $\Delta Q_{ij}=k_i l_j+k_j l_i$.
After squaring and averaging over the direction $\bn$, we finally find
\begin{equation}
    |\mathcal{M}|^2=\frac{g_N^2}{m_N^4 \omega^2}|\sum_\mathcal{O}(V_{\mathcal{O}}(\bk) \overline{u}_3\mathcal{O}u_1\overline{u}_4\mathcal{O}u_2-V_{\mathcal{O}}(\bl) \overline{u}_4\mathcal{O}u_1\overline{u}_3\mathcal{O}u_2)|^2\frac{2 \Delta Q_{ij}\Delta Q_{ij}}{15}.
\end{equation}
This expression should be summed over all the spin states of the incoming and outgoing particles. Noting that $\bk\cdot \bl$ is of order $\omega/m_N$ smaller than $k l$ (see Eq.~\eqref{eq:angle_kl}), we can finally write
\begin{equation}
    |\mathcal{M}|^2=\frac{g_N^2}{m_N^4 \omega^2}|\sum_\mathcal{O}(V_{\mathcal{O}}(\bk) \overline{u}_3\mathcal{O}u_1\overline{u}_4\mathcal{O}u_2-V_{\mathcal{O}}(\bl) \overline{u}_4\mathcal{O}u_1\overline{u}_3\mathcal{O}u_2)|^2\frac{4 k^2 l^2}{15}.
\end{equation}

For $np$ scattering, the expansion must be separately treated for the terms depending on $g$ and on $\delta g$. In squaring the matrix element, the two terms do not interfere, since they lead separately to quadrupole and dipole emission respectively, so their interference vanishes upon angular average. Therefore, we find that the squared matrix element, already averaged over $\bn$, is $|\mathcal{M}|^2=|\mathcal{M}_g|^2+|\mathcal{M}_{\delta g}|^2$, with
\begin{equation}
    |\mathcal{M}_g|^2=\frac{g^2}{m_N^4 \omega^2}|\sum_\mathcal{O}(V_{\mathcal{O}}(\bk) \overline{u}_3\mathcal{O}u_1\overline{u}_4\mathcal{O}u_2-V_{\mathcal{O}}(\bl) \overline{u}_4\mathcal{O}u_1\overline{u}_3\mathcal{O}u_2)|^2\frac{4 k^2 l^2}{15}.
\end{equation}
and
\begin{equation}
    |\mathcal{M}_{\delta g}|^2=\frac{\delta g^2}{m_N^2 \omega^2 } \left|\sum_\mathcal{O}(V_{\mathcal{O}}(\bk) \overline{u}_3\mathcal{O}u_1\overline{u}_4\mathcal{O}u_2-V_{\mathcal{O}}(\bl) \overline{u}_4\mathcal{O}u_1\overline{u}_3\mathcal{O}u_2)\right|^2 \frac{|\mathbf{j}|^2}{3}.
\end{equation}
Notice that $|\mathbf{j}|^2=4k^2$; the asymmetry between $\bk$ and $\bl$ is here induced by the $np$ asymmetry. 

Using the parameterization of Ref.~\cite{Bottaro:2024ugp}, the components of the potential can be identified as $V(\bk)=f$, $V_{\tau}(\bk)=f'$, $V_{\sigma}(\bk)=g$, $V_{\tau\sigma}(\bk)=g'_\bk$, $V_{T\tau}(\bk)=h'_\bk$. The fundamental amplitude to determine is
\begin{equation}
    \mathcal{A}(\bk,\bl)=\sum_\sigma\left|\sum_\mathcal{O}(V_{\mathcal{O}}(\bk) \overline{u}_3\mathcal{O}u_1\overline{u}_4\mathcal{O}u_2-V_{\mathcal{O}}(\bl) \overline{u}_4\mathcal{O}u_1\overline{u}_3\mathcal{O}u_2)\right|^2,
\end{equation}
where the $\sum_\sigma$ denotes a sum over the spins of all incoming and outgoing particles. For $pp$ and $nn$ scattering, we find
\begin{eqnarray}
    &&\mathcal{A}(\bk,\bl)=4(f+f')^2+36 g^2+12\left[g_{\bk}^{'2}+g_{\bl}^{'2}+g'_{\bk} g'_{\bl}+3g(g'_{\bk}+g'_{\bl})\right]-24 g (f+f') -12(f+f')(g'_\bk+g'_\bl)\\ \nonumber &&-4(f+f')(h'_\bk+h'_\bl)+12(h'_\bk+h'_\bl)g+4(h^{'2}_\bk+h^{'2}_\bl)+\frac{4}{3}h'_\bk h'_\bl+4(h'_\bk g'_\bl+h'_\bl g'_\bk+2h'_\bk g'_\bk+2h'_\bl g'_\bl),
\end{eqnarray}
while for $np$ scattering we have (with the convention that $\bk=\bp_3-\bp_1$ is the momentum transfer between the two neutrons and $\bl=\bp_4-\bp_1$ is the momentum transfer between neutron and proton)
\begin{eqnarray}
    && \mathcal{A}(\bk,\bl)=4f^2+28 f^{'2}+36g^2+12 g^{'2}_\bk+48 g^{'2}_\bl +4f f'-24 f g-12 f' g +12 f g'_\bk-24 f g'_\bl+24 f' (g'_\bk+g'_\bl)\\ \nonumber &&-36 g g'_\bk+72 g g'_\bl-24 g'_\bk g'_\bl
    +4h'_\bk(f+2f'-3g+2g'_\bk-2g'_\bl)-8h'_\bl(f-f'-3g+g'_\bk-4g'_\bl)-\frac{8}{3}h'_\bk h'_\bl+4h^{'2}_\bk+16 h^{'2}_{\bl}.
\end{eqnarray}

\subsection{Explicit integration for $nn$ and $pp$ scattering}

For $nn$ and $pp$ scattering, after replacing the expressions from Ref.~\cite{Bottaro:2024ugp} and introducing the notation
\begin{equation}
    D^k_{\pi}=\frac{\tk^2}{\tk^2+\alpha_\pi},\;D^l_{\pi}=\frac{\tl^2}{\tl^2+\alpha_\pi},\;D^k_{\rho}=\frac{\tk^2}{\tk^2+\alpha_\rho},\;D^l_{\rho}=\frac{\tl^2}{\tl^2+\alpha_\rho},
\end{equation}
where $\alpha_{\pi}=m_\pi^2/M T$ and similarly for $\alpha_\rho$
, we can expand the squared amplitude as
\begin{eqnarray}\label{eq:expansion_amplitude_nn}
    &&\mathcal{A}=C_0+\sum_{i=k,l}\sum_{\mu=\pi,\rho} C^i_\mu D^i_\mu+\sum_{i,j=k,l}\sum_{\mu,\nu=\pi,\rho} C^{ij}_{\mu\nu} D^{i}_{\mu} D^j_\nu,
\end{eqnarray}
with the obvious identity $C^{ij}_{\mu\nu}=C^{ji}_{\nu\mu}$.
The coefficients of this expansion can be found explicitly
\begin{eqnarray}
    &&C_0=4\left[(f+f')^2-6(f+f')(g+g')+9(g+g')^2\right],\\
    \nonumber && C^k_\pi=C^l_\pi=\frac{4f_\pi^2}{m_\pi^2}(f+f'-3g-3g'),\; C^k_\rho=C^l_\rho=\frac{8 C_\rho f_\pi^2}{m_\pi^2}(f+f'-3g-3g'),\\ \nonumber &&C^{kk}_{\pi\pi}=\frac{4f_\pi^4}{m_\pi^4},\; C^{kk}_{\rho\rho}=\frac{8C_\rho^2 f_\pi^4}{m_\pi^4},\; C^{kl}_{\pi\pi}=C^{lk}_{\pi\pi}=\frac{1}{2}\frac{4f_\pi^4}{3m_\pi^4},\; C^{kl}_{\rho\rho}=C^{lk}_{\rho\rho}=\frac{1}{2}\frac{16 C_\rho^2 f_\pi^4}{3m_\pi^4},\\ \nonumber && C^{kl}_{\pi \rho}=C^{lk}_{\pi\rho}=C^{kl}_{\rho\pi}=C^{lk}_{\rho\pi}=\frac{1}{2}\frac{8 C_\rho f_\pi^4}{3 m_\pi^4};
\end{eqnarray}
the factors $1/2$ are explicitly shown as a reminder of the fact that the corresponding integrand function must be counted twice since $C^{ij}_{\mu\nu}=C^{ji}_{\nu\mu}$. 

With this expansion, we can identify a limited number of integrals that need to be performed
\begin{eqnarray}
    &&I_0=\int \frac{d\tk d\tl}{\sqrt{4\tp_F^2-\tk^2-\tl^2}} \tk^2 \tl^2,\; I_k(\alpha)=\int \frac{d\tk d\tl}{\sqrt{4\tp_F^2-\tk^2-\tl^2}} \tk^2 \tl^2 \frac{\tk^2}{(\tk^2+\alpha)},\\ \nonumber && I_l(\alpha)=\int \frac{d\tk d\tl}{\sqrt{4\tp_F^2-\tk^2-\tl^2}} \tk^2 \tl^2 \frac{\tl^2}{(\tl^2+\alpha)},
    \; I_{kk}(\alpha,\beta)=\int  \frac{d\tk d\tl}{\sqrt{4\tp_F^2-\tk^2-\tl^2}} \tk^2 \tl^2 \frac{\tk^4}{(\tk^2+\alpha)(\tk^2+\beta)},\\ \nonumber &&
    I_{ll}(\alpha,\beta)=\int  \frac{d\tk d\tl}{\sqrt{4\tp_F^2-\tk^2-\tl^2}} \tk^2 \tl^2 \frac{\tl^4}{(\tl^2+\alpha)(\tl^2+\beta)}, I_{kl}(\alpha,\beta)=\int \frac{d\tk d\tl}{\sqrt{4\tp_F^2-\tk^2-\tl^2}} \tk^2 \tl^2 \frac{\tk^2 \tl^2}{(\tk^2+\alpha)(\tl^2+\beta)}.
\end{eqnarray}

All these integrals can be explicitly performed to give
\begin{eqnarray}
&&I_0=\frac{16\pi \tp_F^5}{15},\\
\nonumber
&& I_k(\alpha)=I_l(\alpha)=\frac{\pi}{60}\left[64\tp_F^5-80\tp_F^3\alpha-30\tp_F \alpha^2+15\alpha^{3/2}(4\tp_F^2+\alpha)\arctan\left(\frac{2\tp_F}{\sqrt{\alpha}}\right)\right],
\\ \nonumber
&&I_{kk}(\alpha,\beta)=I_{ll}(\alpha,\beta)=\frac{\pi}{60(\alpha-\beta)}\left[64 \tp_F^5(\alpha-\beta)+80\tp_F^3 (\beta^2-\alpha^2)+30 \tp_F (\beta^3-\alpha^3)\right.\\ \nonumber &&\left.+15\alpha^{5/2}(4\tp_F^2+\alpha) \arctan\left(\frac{2\tp_F}{\sqrt{\alpha}}\right)-15\beta^{5/2}(4\tp_F^2+\beta) \arctan\left(\frac{2\tp_F}{\sqrt{\beta}}\right)\right],\\
\nonumber
&&I_{kl}(\alpha,\beta)=\frac{\pi}{60}\left[64 \tp_F^5-30 \tp_F(\alpha-\beta)^2-80 \tp_F^3(\alpha+\beta)\right.\\ \nonumber &&\left.+15\alpha^{3/2} (4\tp_F^2+\alpha-2\beta)\arctan\left(\frac{2\tp_F}{\sqrt{\alpha}}\right)+15\beta^{3/2} (4\tp_F^2+\beta-2\alpha)\arctan\left(\frac{2\tp_F}{\sqrt{\beta}}\right)\right.\\ \nonumber &&\left.+\frac{30\alpha^{3/2} \beta^{3/2}}{\sqrt{4\tp_F^2+\alpha+\beta}}\arctan\left(\frac{2\tp_F \sqrt{4\tp_F^2+\alpha+\beta}}{\sqrt{\alpha\beta}}\right)\right],
\end{eqnarray}
with $\tp_F=p_F/\sqrt{m_NT}$. In terms of these integrals, we finally have
\begin{equation}
    \mathcal{N}_S=\frac{|{\bf q}| m_N (m_N T)^{7/2} J(\tilde{\omega})}{32\pi^8}\frac{4 g_N^2 m_N^2 T^2}{15 m_N^4 \omega^2}\left[C_0 I_0+\sum_{i=k,l}\sum_{\mu=\pi,\rho} C^i_\mu I_i(\alpha_\mu)+\sum_{i,j=k,l} \sum_{\mu,\nu=\pi,\rho} C^{ij}_{\mu\nu} I_{ij}(\alpha_\mu,\alpha_\nu)\right].
\end{equation}

Starting from this expression, the scalar emissivity from $nn$ and $pp$ channels reads

\begin{equation}\label{eq:final_nn}
    \begin{split}
        Q^S_{NN}&= \int_0^\infty\,d\omega\,\omega \,\mathcal{N}_S\\
        &=g_N^2\,\frac{\,m_N^{5/2}\, T^{13/2}}{120\,\pi^8}\, B(\tm) \,\left[C_0 I_0+\sum_{i=k,l}\sum_{\mu=\pi,\rho} C^i_\mu I_i(\alpha_\mu)+\sum_{i,j=k,l} \sum_{\mu,\nu=\pi,\rho} C^{ij}_{\mu\nu} I_{ij}(\alpha_\mu,\alpha_\nu)\right]\,.
    \end{split}
\end{equation}

where
\begin{equation}
    B(\tm)=\int_{\tilde{m}_\phi}^{\infty}\,d\tom\, \tom \sqrt{\tom^2-\tilde{m}_\phi^2}\,\frac{J(\tom)}{\tom^2}\simeq\frac{11}{90}\pi^4\,e^{-0.7606\,\tm}\,. 
\end{equation}

\subsection{Explicit integration for $np$ scattering}

In the case of $np$ scattering, we can make use of the simplification, that was validated in Ref.~\cite{Bottaro:2024ugp} in the case of neutrino emission, that $\tp_p\ll \tp_n$. From Eq.~\eqref{eq:N_a_final}, we see that the integration region for $\tl$ is restricted to $\tp_n-\tp_p<\tl<\tp_n+\tp_p$. Therefore, with good approximation, in the matrix element we can simply take $\tl=\tp_n$ (essentially the exchanged momentum between protons and neutrons is constant and equal to the much larger neutron momentum). The integral over $\tl$ in Eq.~\eqref{eq:N_a_final} simply leads to
\begin{equation}
    \mathcal{N}_S=\frac{|{\bf q}| m_N (m_NT)^{7/2} J(\tilde{\omega})}{64\pi^7}\frac{4m_N T}{3\omega^2 m_N^2}\left(\delta g^2 + \frac{g^2 p_N^2}{m_N^2}\right)\int_0^{2\tp_p} d\tk \tk^2 \mathcal{A},
\end{equation}
where the integral over $\tk$ is performed up to the maximum value that renders the argument of the square root in Eq.~\eqref{eq:N_a_final} positive for $\tl=\tp_n$. Physically, since $\tk$ is the momentum exchange between the two protons, it cannot exceed $2\tp_p$.

It is convenient to use the same expansion as in the $nn$ and $pp$ scattering in Eq.~\eqref{eq:expansion_amplitude_nn}; in this case the coefficients of the expansion are
\begin{eqnarray}
    &&C_0=4\left[f^2+7f^{'2}-3f'(g-4g')+9(g^2+g g'+g^{'2})+f(f'-3(2g+g'))\right],\\ 
    \nonumber && C^k_\rho=2 C_\rho C^k_\pi=\frac{8C_\rho f_\pi^2 (3g-f-2f')}{m_\pi^2},\; C^l_\rho=2 C_\rho C^l_\pi=\frac{16 C_\rho f_\pi^2 (f-f'-3(g+g'))}{m_\pi^2},\\ \nonumber &&
    C^{ll}_{\pi\pi}=4C^{kk}_{\pi\pi}=-12 C^{kl}_{\pi\pi}=\frac{16 f_\pi^4}{m_\pi^4},\; C^{ll}_{\rho\rho}=4 C^{kk}_{\rho\rho}=-6C^{kl}_{\rho\rho}=\frac{32 C_\rho^2 f_\pi^4}{m_\pi^4},\\ \nonumber && C^{kl}_{\pi \rho}=C^{lk}_{\pi\rho}=C^{kl}_{\rho\pi}=C^{lk}_{\rho\pi}=-\frac{1}{2}\frac{16 C_\rho f_\pi^4}{3 m_\pi^4}.
\end{eqnarray}

In the expansion, we encounter three fundamental integrals
\begin{equation}
    I_0=\int_0^{2\tp_p} d\tk \tk^2, \; I_k(\alpha)=\int_0^{2\tp_p}d\tk \tk^2 \frac{\tk^2}{\tk^2+\alpha},\; I_{kk}(\alpha,\beta)=\int_0^{2\tp_p}d\tk \tk^2 \frac{\tk^4}{(\tk^2+\alpha)(\tk^2+\beta)};
\end{equation}
all of these are elementary and yield
\begin{eqnarray}
    &&I_0=\frac{8\tp_p^3}{3},\\ \nonumber &&I_k(\alpha)=\frac{8\tp_p^3}{3}-2\tp_p \alpha+\alpha^{3/2}\arctan\left(\frac{2\tp_p}{\sqrt{\alpha}}\right),\\ \nonumber
    &&I_{kk}(\alpha,\beta)=\frac{2\tp_p(\alpha-\beta)\left[4\tp_p^2-3(\alpha+\beta)\right]+3\alpha^{5/2}\arctan\left(\frac{2\tp_p}{\sqrt{\alpha}}\right)-3\beta^{5/2}\arctan\left(\frac{2\tp_p}{\sqrt{\beta}}\right)}{3(\alpha-\beta)}.
\end{eqnarray}

Therefore, in integrating each of the terms in the expansion associated with the coefficients $C_0$, $C^i_\mu$, $C^{ij}_{\mu\nu}$, we encounter the following fundamental integrals
\begin{eqnarray}
    &&I^k_\mu=I_k(\alpha_\mu),\; I^l_\mu=I_0\frac{\tp_n^2}{\tp_n^2+m_\mu^2},\; I^{kk}_{\mu\nu}=I_{kk}(\alpha_\mu,\alpha_\nu),\; I^{ll}_{\mu\nu}=I_0 \frac{\tp_n^4}{(\tp_n^2+\alpha_\mu)(\tp_n^2+\alpha_\nu)},\\ \nonumber && I^{kl}_{\mu\nu}=I^{lk}_{\nu\mu}= I_k(\alpha_\mu) \frac{\tp_n^2}{\tp_n^2+\alpha_\nu}. 
\end{eqnarray}
With these definitions, we finally get
\begin{equation}\label{eq:scalar_emission_np}
     \mathcal{N}_s=\frac{|{\bf q}| m_N (m_NT)^{7/2} J(\tilde{\omega})}{64\pi^7}\frac{4m_N T}{3\omega^2 m_N^2}\left(\delta g^2 + \frac{g^2 p_n^2}{5M^2}\right)\left[C_0 I_0+\sum_{i=k,l} \sum_{\mu=\pi,\rho} C^i_\mu I^i_\mu+\sum_{i,j=k,l} \sum_{\mu,\nu=\pi,\rho} C^{ij}_{\mu\nu} I^{ij}_{\mu\nu}\right].
\end{equation}

Then, the scalar emissivity from the $np$ channel is given by

\begin{equation}\label{eq:final_np}
    \begin{split}
        Q^S_{np}&= \int_0^\infty\,d\omega\,\omega \,\mathcal{N}_s\\
        &=\left(\delta g^2 +\frac{p_n^2}{5m_N^2}\,g^2\right)\,\frac{m_N^{7/2}\, T^{11/2}}{48\,\pi^7}\, B(\tm)\,\left[C_0 I_0+\sum_{i=k,l}\sum_{\mu=\pi,\rho} C^i_\mu I_i(\alpha_\mu)+\sum_{i,j=k,l} \sum_{\mu,\nu=\pi,\rho} C^{ij}_{\mu\nu} I_{ij}(\alpha_\mu,\alpha_\nu)\right]\,.
    \end{split}
\end{equation}

Therefore, overall we use Eqs.~\eqref{eq:final_nn} and~\eqref{eq:final_np} to determine the emissivities from $nn$ (and $pp$) and $np$ bremsstrahlung respectively.

\end{widetext}

\section{B.~Neutron star observations}
In this work we employ data referring to five isolated NSs with ages $\sim10^{5}\,$yrs, for which thermal luminosity and kinematic data are available. Estimations for the NS thermal luminosities within their uncertainties are inferred on the basis of related X-ray observations interpreted in light of NS cooling theory~\cite{Potekhin:2020ttj}, while quantitative estimations for their ages are obtained by analyzing the kinematic displacement of the remnant from the location of the supernova event giving birth to the pulsar~\cite{Suzuki:2021ium}. In analogy to Ref.~\cite{Buschmann:2021juv}, we consider NSs in the same cooling epoch, in which light scalar emissivities ($\propto T^{4}$) is expected to be the dominant cooling channel compared to neutrino emissivity~($\propto T^{8}$). Thus, NSs with these ages are sensitive probes to constrain the properties of such particles. In this regard, we highlight that we always refer to measurements of the total thermal luminosity, since it is a more robust observable compared to surface temperature, which is affected by local inhomogeneities induced by the strong surrounding magnetic fields. Moreover, Gaussian priors on the age and luminosity measurements will be assumed.

The relevant data for our analysis are listed in the main text in Tab.~I. J1856 and J1308 have originated in the Upper Scorpius OB~\cite{Motch:2009nq, Mignani:2012mm} and their luminosity data are inferred on the basis of a NS atmosphere model with a thin layer of partially-ionized hydrogen or double black body spectrum. These two approaches applied on J1856 suggest a lower luminosity bound at $L_\gamma\simeq5\times10^{31}\,\erg/\s$ and an upper bound at $L_\gamma\simeq8\times10^{31}\,\erg/\s$. For J1308, the same models suggest $L_\gamma\simeq(3.3\pm0.5)\times10^{32}\,\erg/\s$ and  $L_\gamma\simeq2.6\times10^{32}\,\erg/\s$, respectively. In the case of J0720, born in the Trumpler association~\cite{Tetzlaff:2011kh}, both models lead to $L_\gamma\simeq2\times10^{32}\,\erg/\s$. The birth of the J1605 can be related to a binary disruption induced by a SN explosion~\cite{Tetzlaff:2012rz}, and its present luminosity is obtained from a double blackbody fit, yielding $L_\gamma\simeq(4\pm1)\times10^{32}\,\erg/\s$. Finally, the pulsar J0659 is located within the large diffuse SN remnant Monogem Ring~\cite{Thorsett:2003xy}. Its emission is fit by a double blackbody spectrum including a broken power-law component to account for hard X-ray pulsed emission from the pulsar magnetosphere~\cite{Zharikov:2021llh}, leading to a total luminosity $L_\gamma\simeq(2.8\pm1.4)\times10^{32}\,\erg/\s$.

\begin{figure}
    \includegraphics[width=1\linewidth]{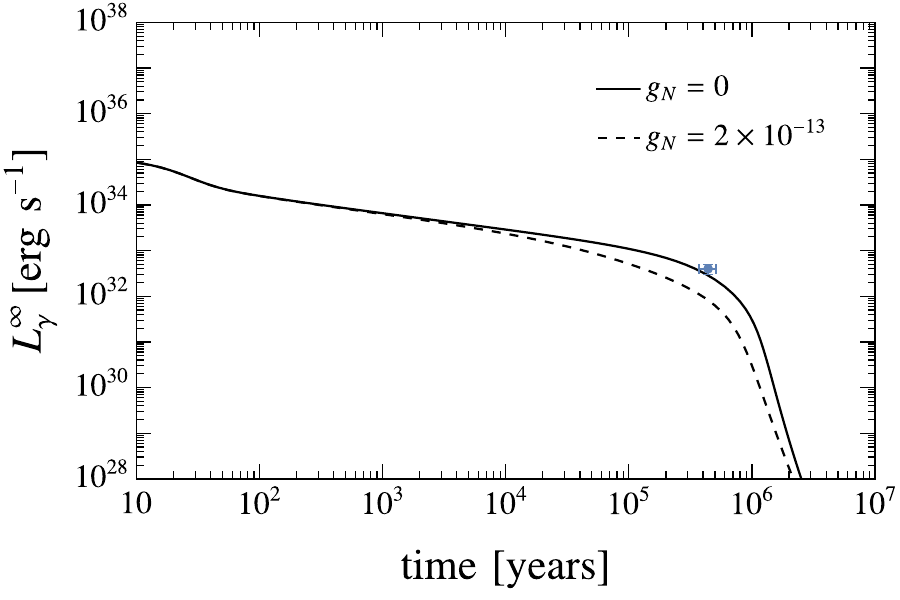}
    \caption{Cooling light-curves derived from the best fit NS model for J1605 assuming no BSM physics. We compare the behavior of the light-curves for the unperturbed NS model ($g_N=0$) and by adding on top of the NS model the emissivity of scalars with coupling strength $g_N=2\times10^{-13}$.}\label{figReport}
\end{figure}

In analogy to Ref.~\cite{Buschmann:2021juv}, we consider NSs in the same cooling epoch, in which light scalar emissivities~($\propto T^4$) is expected to be the dominant cooling channel compared to neutrino emissivity ($\propto T^8$). Thus, the introduction of an exotic energy-loss channel via light-scalar may substantially accelerate the standard cooling process of isolated neutron stars, leading to tension between the observed and theoretically predicted luminosities. This effect is clearly visible in Fig.~\ref{figReport}, showing that the inclusion of an efficient scalar cooling channel with coupling $g_N=2\times10^{-13}$---chosen to magnify the effect on such a plot-- would significantly reduce the photon luminosity at $\sim10^{5}$ yrs compared to the best-fit NS model for J1605 assuming no BSM physics~($g_N=0$).

\section{C.~Comparison with other bounds on scalars and axions}
\begin{figure*}
    \centering
    \includegraphics[width=0.95\textwidth]{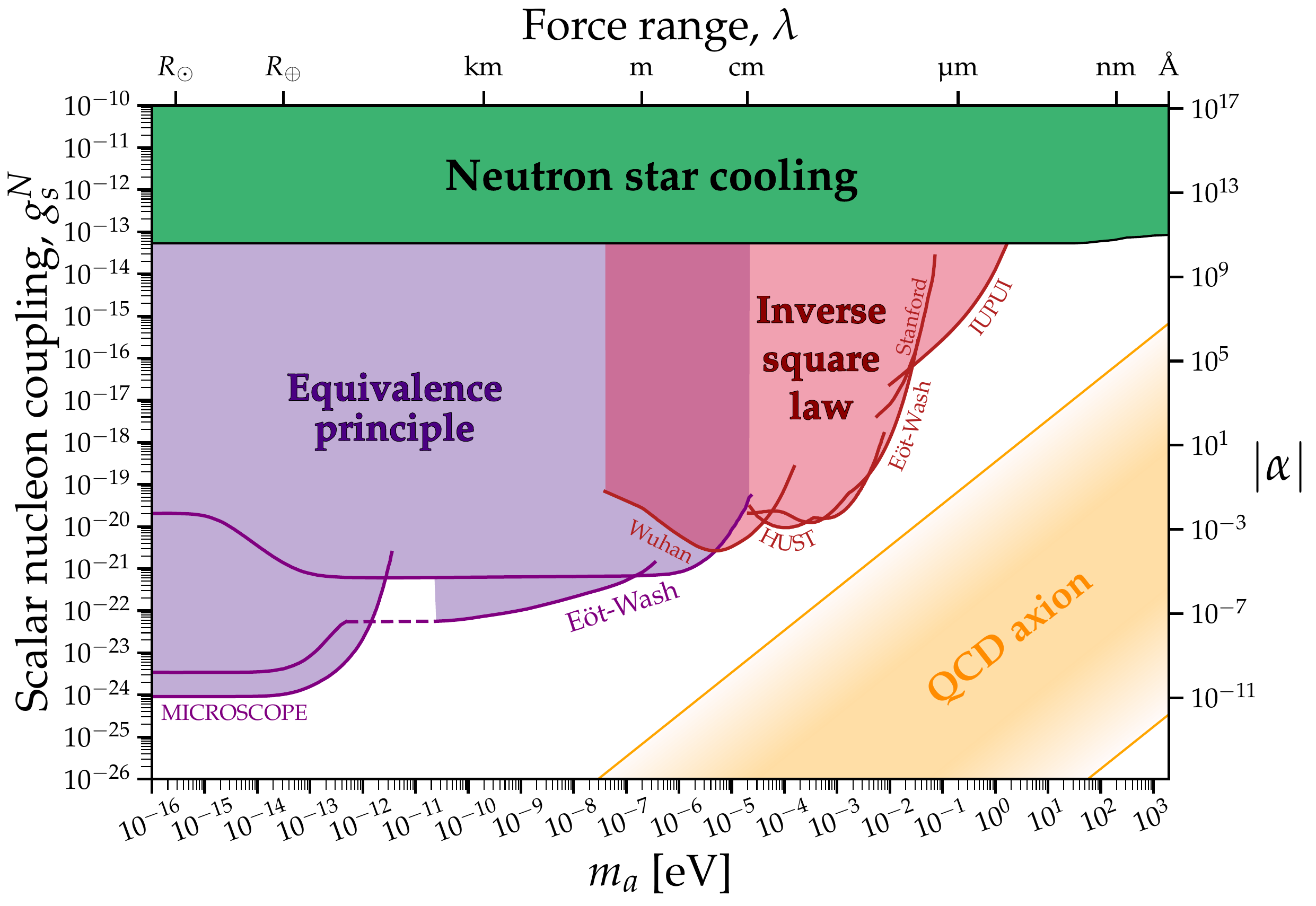}
    \caption{Full landscape of constraints on the scalar-nucleon coupling for scalar masses smaller than $\sim$keV. The bound from this work is shown in green. Laboratory constraints from tests of the violation of the inverse square law and the weak equivalence principle are shown in red and purple respectively. The references for these bounds are as follows: MICROSCOPE~\cite{Berge:2017ovy,Berge:2021yye,MICROSCOPE:2022doy}, E\"ot-Wash~\cite{Smith:1999cr,Kapner:2006si,Lee:2020zjt}, Wuhan~\cite{Ke:2021jtj}, HUST~\cite{Tu:2007zz,Yang:2012zzb,Tan:2016vwu, Tan:2020vpf}, IUPUI~\cite{Chen:2014oda}.}
    \label{fig:ScalarNucleon}
\end{figure*}

\begin{figure*}
    \centering
    \includegraphics[width=0.95\textwidth]{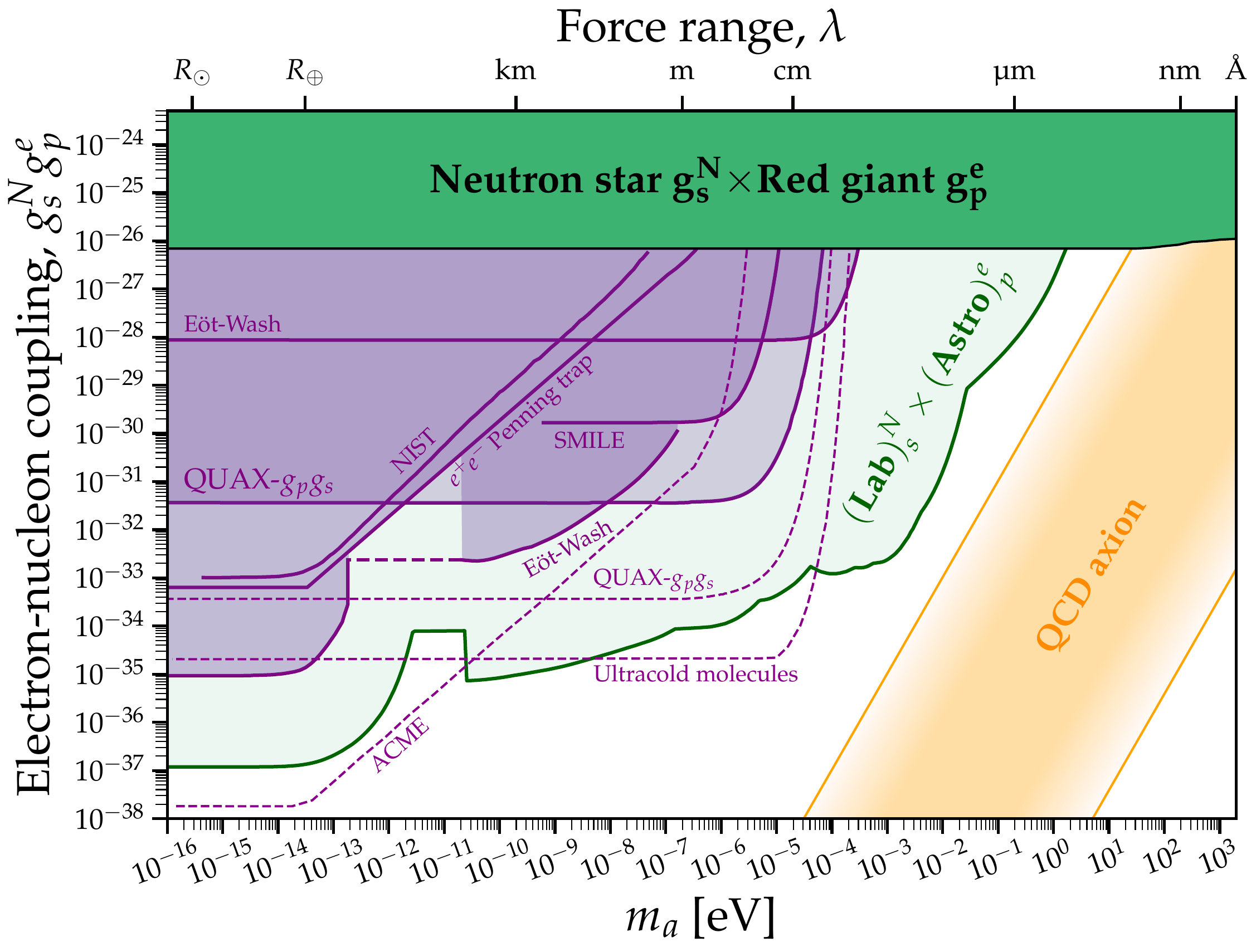}
    \caption{Limits on the scalar coupling to nucleons multiplied by the pseudoscalar coupling to electrons. We show the combined astrophysical bound on this coupling combination in green, which is derived by multiplying the neutron star cooling bound from this work with the tip-of-the-red-giant branch bound from Ref.~\cite{Capozzi:2020cbu}. Existing laboratory bounds are shown in purple (E\"ot-Wash~\cite{Heckel:2008hw}, QUAX~\cite{Crescini:2017uxs}, NIST~\cite{Wineland:1991zz}, $e^+ e^-$ penning trap~\cite{Fan:2023hci}, SMILE~\cite{Lee:2018vaq}), while future projections are shown as dashed lines (ACME~\cite{Agrawal:2023lmw}, ultracold molecules~\cite{Agrawal:2023lmw}, QUAX~\cite{Crescini:2016lwj}). The combined Lab $\times$ Astro bound is derived by multiplying the laboratory bound on the scalar-nucleon coupling from Fig.~\ref{fig:ScalarNucleon} with the red giant bound on pseudoscalars~\cite{Capozzi:2020cbu}.}
    \label{fig:MonopoleDipole_ElectronNucleon}
\end{figure*}

\begin{figure*}
    \centering
    \includegraphics[width=0.95\textwidth]{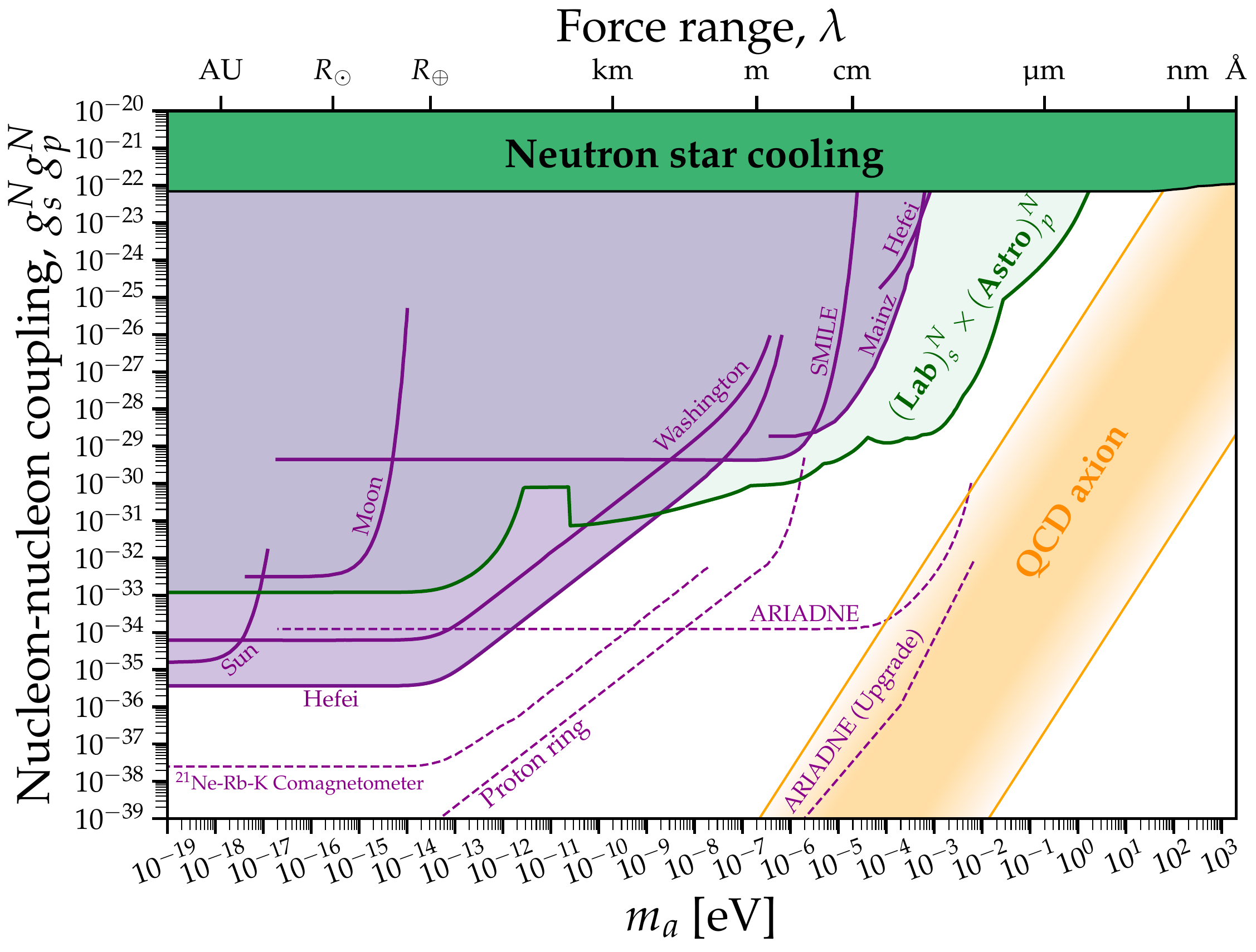}
    \caption{Limits on the scalar coupling to nucleons multiplied by the pseudoscalar coupling to nucleons. We show the combined neutron star cooling bound from this work and Ref.~\cite{Buschmann:2021juv} in green. Existing laboratory bounds are shown in purple (Sun~\cite{Wu:2023ypz}, Moon~\cite{Wu:2023ypz}, Washington~\cite{Venema:1992zz}, SMILE~\cite{Lee:2018vaq}, Mainz~\cite{Tullney:2013wqa} and Hefei~\cite{Feng:2022tsu}), while future projections as dashed lines ($^{21}$Ne-Rb-K comagnetometer~\cite{Wei:2022ggs}, proton ring~\cite{Agrawal:2022wjm} and ARIADNE~\cite{Arvanitaki:2014dfa, Geraci:2017bmq}). The combined Lab $\times$ Astro bound is derived by multiplying the laboratory bound on the scalar-nucleon coupling from Fig.~\ref{fig:ScalarNucleon} with the neutron star cooling bound on pseudoscalars~\cite{Buschmann:2021juv}.}
    \label{fig:MonopoleDipole_NucleonNucleon}
\end{figure*}

As we have shown in the main body of this article, astrophysical bounds are the most stringent on fifth forces that have a range smaller than a few microns, equivalent to scalar masses $\gtrsim$eV. However, the emission rate of exotic species is typically insensitive to the value of the particle's mass if it is much smaller than the temperature of the environment from which the particle is produced. So laboratory tests are expected to take over for $m_\phi \lesssim{\rm eV}$. In the case of scalars with equal nucleon couplings, the shortest-range laboratory tests to surpass our new bound from neutron star cooling are those performing tests of violations to the Newtonian inverse square law. In the mass range shown in Fig.~1 of the main article, the leading constraint spanning the range $\lambda =30$--$8000$~nm comes from a differential force
measurement using a microelectromechanical torsional oscillator at the Indiana University–Purdue
University Indianapolis (IUPUI)~\cite{Chen:2014oda}. The challenging background due to vacuum fluctuations makes experimental progress very challenging in this regime. The result from Ref.~\cite{Chen:2014oda} was only possible thanks to a novel technique in which their source mass was coated with a gold film thicker than the material’s plasma wavelength, which acts to suppress the Casimir force between the interior of the source mass and the attractor. 

It is possible to experimentally test for fifth forces acting over shorter ranges than this. Experiments using levitated test masses~\cite{Blakemore:2021zna,Venugopalan:2024kgu}, or similar setups to those used to measure the Casimir effect~\cite{Chiu:2009fqu,Bezerra:2010pq,Sushkov:2011md,Banishev:2012kkb,Banishev:2014jka,Mostepanenko:2020lqe,Klimchitskaya:2013rwd,Klimchitskaya:2023cgy} are relevant at the sub-micron scale; while neutron scattering experiments~\cite{Pokotilovski:2006up,Haddock:2017wav,Heacock:2021btd,Kamiya:2015eva,Bogorad:2023zmy} represent the most sensitive technique for sub-nm scales. The level of current constraints from these approaches is $\alpha \sim 10^{11}$--$10^{22}$ between $\lambda \sim 100$--$0.1$~nm (see e.g.~Ref.~\cite{Klimchitskaya:2023cgy} for the most recent summary), which is many orders of magnitude higher than even the weaker of our astrophysical bounds. 

To provide more context for our main figure, we also present here a summary of bounds on lower-mass scalars in Fig.~\ref{fig:ScalarNucleon}. At ranges $10~{\rm m}~\gtrsim\lambda \gtrsim\upmu$m, tests of the inverse square law between two masses have leading sensitivity, while tests of the equivalence principle using torsion balances and accelerometers dominate for much longer-range forces. This plot updates previous summary plots of this parameter space presented in Refs.~\cite{Raffelt:2012sp,OHare:2020wah,AxionLimits}. We refer to the figure caption for a list of references for each bound.

An interesting ramification of the new bound on scalars we have derived in this work is that it advances bounds on the axion as well. Although the axion is a pseudoscalar particle, it is believed that it may possess scalar couplings
to fermions, which break CP, in addition to its usual CP-preserving pseudoscalar couplings~\cite{Moody:1984ba, Georgi:1986kr, Pospelov:1997uv, Pospelov:2005pr,Plakkot:2023pui,Dekens:2022gha,DiLuzio:2023cuk}. It is possible to assign a range of values of $g_s^N$ expected for QCD axion models by estimating how much the QCD $\bar{\theta}$ angle may be shifted away from the CP-preserving value of zero. This has been discussed previously in e.g.~Refs.~\cite{Bigazzi:2019hav,Bertolini:2020hjc,DiLuzio:2023lmd}, as well as recently in Ref.~\cite{DiLuzio:2024ctr} in the broader context of alternative axion models. A ``QCD axion band'' for scalar couplings---inspired by equivalent bands for other couplings---can then be bounded from above by the current experimental upper limit on $\bar{\theta}$ from the absence of the neutron's electric dipole moment~\cite{Abel:2020pzs}, and from below by the expected level of CP violation induced through the weak interaction, quantified by the Jarlskog invariant~\cite{Georgi:1986kr,Ellis:1978hq,Khriplovich:1985jr,Gerard:2012ud,Okawa:2021fto}.

If axions possess such scalar couplings to fermions, then this opens up a novel angle in which to test for the existence of axions---either as a mediator of new forces, or as dark matter---in a way that is complementary to other probes which rely only on their derivative or pseudoscalar couplings. In particular, the combined probes of, say, $g_s^N g_p^N$ for a nucleon-nucleon scalar-pseudoscalar interaction mediated by an axion, or $g_s^N g_p^e$ for a nucleon-electron interaction, are potentially advantageous routes to search for the axion because the corresponding force is spin-dependent but only requires one spin-polarised sample rather than the two needed to search for dipole-dipole forces. Searches for new monopole-dipole interactions between nucleons and/or electrons of this kind have been performed over decades by many experimental groups; we refer to Ref.~\cite{Cong:2024qly} for a review.

However, as was first pointed out in Ref.~\cite{Raffelt:2012sp} (and updated in Ref.~\cite{OHare:2020wah}), direct experimental tests of such monopole-dipole forces at short ranges need to compete with the enormously stringent combination of the scalar and pseudoscalar bounds from stellar cooling. We illustrate this in Figs.~\ref{fig:MonopoleDipole_ElectronNucleon} and~\ref{fig:MonopoleDipole_NucleonNucleon}, which updates similar summary plots presented in Refs.~\cite{OHare:2020wah,AxionLimits}. 

Figure~\ref{fig:MonopoleDipole_ElectronNucleon}, first of all, shows the landscape of constraints on the coupling combination $g_s^N g_p^e$ which can be tested experimentally earching for a macroscopic monopole-dipole force between spin-polarised electrons and a source mass containing nucleons. A dedicated experiment as part of the QUAX program is searching for such forces---a future projection is also shown on the plot alongside several other proposals. So far, none of the existing constraints can compete with the product of the laboratory bounds on $g_s^N$ and the leading astrophysical bound on $g_p^e$, which in this case comes from the luminosity of the tip of the red giant branch in the $\omega$Cen globular cluster~\cite{Capozzi:2020cbu}. The pure astrophysical bound $g_s^N g_p^e \sim 7 \times 10^{-27}$ shown in dark green, is found by combining the neutron star cooling bound (this work) with the latter red giant bound---this remains the leading bound up to masses $\sim$100~keV, after which the combination of our neutron star bound on scalars and SNe bounds on pseudoscalars~\cite{Carenza:2021pcm,Ferreira:2022xlw,Fiorillo:2025sln} will take over. Other laboratory searches at these higher masses, for example using atomic and molecular systems e.g.~\cite{Stadnik:2017hpa,Baruch:2024fbh,Baruch:2024frj,Liu:2025ows}---are typically much weaker than the astrophysical bounds.

Lastly, Fig.~\ref{fig:MonopoleDipole_NucleonNucleon} shows the equivalent plot to the previous one but for the combination $g_s^N g_p^N$. This coupling can be tested experimentally by searching for a monopole-dipole force between nucleons. In this case, the pure astrophysical bound is $g_s^N g_p^n \sim 7 \times 10^{-23}$ which originates on both sides from neutron star cooling (this work and Ref.~\cite{Buschmann:2021juv}). Unlike the previous case, nucleon-nucleon force searches at very long distances do outperform the combined laboratory and astrophysical bound, corresponding to masses $\lesssim 10^{-9}$~eV.

In light of these updates, we encourage experimental groups searching for new forces to keep the strong astrophysical bounds in mind, particularly those at large masses/short ranges. Although it is possible to engineer models in which astrophysical bounds are relaxed compared to laboratory bounds, e.g.~by suppressing the emission of light new particles from stellar environments~\cite{Jain:2005nh, Masso:2005ym, Jaeckel:2006xm, Masso:2006gc, Budnik:2020nwz, DeRocco:2020xdt, Bloch:2020uzh}, inspired by  chameleon models (see e.g. Ref.~\cite{Khoury:2003aq}), such constructions are generally not especially simple or natural.



%% file: references.bib
@article{Sushkov:2011md,
    author = "Sushkov, A. O. and Kim, W. J. and Dalvit, D. A. R. and Lamoreaux, S. K.",
    title = "{New Experimental Limits on Non-Newtonian Forces in the Micrometer Range}",
    eprint = "1108.2547",
    archivePrefix = "arXiv",
    primaryClass = "quant-ph",
    doi = "10.1103/PhysRevLett.107.171101",
    journal = "Phys. Rev. Lett.",
    volume = "107",
    pages = "171101",
    year = "2011"
}

@article{Khoury:2003aq,
    author = "Khoury, Justin and Weltman, Amanda",
    title = "{Chameleon fields: Awaiting surprises for tests of gravity in space}",
    eprint = "astro-ph/0309300",
    archivePrefix = "arXiv",
    doi = "10.1103/PhysRevLett.93.171104",
    journal = "Phys. Rev. Lett.",
    volume = "93",
    pages = "171104",
    year = "2004"
}

@article{Arvanitaki:2017nhi,
    author = "Arvanitaki, Asimina and Dimopoulos, Savas and Van Tilburg, Ken",
    title = "{Resonant absorption of bosonic dark matter in molecules}",
    eprint = "1709.05354",
    archivePrefix = "arXiv",
    primaryClass = "hep-ph",
    doi = "10.1103/PhysRevX.8.041001",
    journal = "Phys. Rev. X",
    volume = "8",
    number = "4",
    pages = "041001",
    year = "2018"
}

@article{Iwamoto:1984ir,
    author = "Iwamoto, N.",
    title = "{Axion Emission from Neutron Stars}",
    doi = "10.1103/PhysRevLett.53.1198",
    journal = "Phys. Rev. Lett.",
    volume = "53",
    pages = "1198--1201",
    year = "1984"
}

@article{Buschmann:2021juv,
    author = "Buschmann, Malte and Dessert, Christopher and Foster, Joshua W. and Long, Andrew J. and Safdi, Benjamin R.",
    title = "{Upper Limit on the QCD Axion Mass from Isolated Neutron Star Cooling}",
    eprint = "2111.09892",
    archivePrefix = "arXiv",
    primaryClass = "hep-ph",
    doi = "10.1103/PhysRevLett.128.091102",
    journal = "Phys. Rev. Lett.",
    volume = "128",
    number = "9",
    pages = "091102",
    year = "2022"
}

@article{Pokotilovski:2006up,
    author = "Pokotilovski, Yu. N.",
    title = "{Constraints on new interactions from neutron scattering experiments}",
    eprint = "hep-ph/0601157",
    archivePrefix = "arXiv",
    doi = "10.1134/S1063778806060020",
    journal = "Phys. Atom. Nucl.",
    volume = "69",
    pages = "924--931",
    year = "2006"
}

@article{Bogorad:2023zmy,
    author = "Bogorad, Zachary and Graham, Peter W. and Gratta, Giorgio",
    title = "{Detecting nanometer-scale new forces with coherent neutron scattering}",
    eprint = "2303.17744",
    archivePrefix = "arXiv",
    primaryClass = "hep-ph",
    reportNumber = "FERMILAB-PUB-23-160-SQMS-V",
    doi = "10.1103/PhysRevD.108.055005",
    journal = "Phys. Rev. D",
    volume = "108",
    number = "5",
    pages = "055005",
    year = "2023"
}

@article{Blakemore:2021zna,
    author = "Blakemore, Charles P. and Fieguth, Alexander and Kawasaki, Akio and Priel, Nadav and Martin, Denzal and Rider, Alexander D. and Wang, Qidong and Gratta, Giorgio",
    title = "{Search for non-Newtonian interactions at micrometer scale with a levitated test mass}",
    eprint = "2102.06848",
    archivePrefix = "arXiv",
    primaryClass = "hep-ex",
    doi = "10.1103/PhysRevD.104.L061101",
    journal = "Phys. Rev. D",
    volume = "104",
    number = "6",
    pages = "L061101",
    year = "2021"
}

@article{Venugopalan:2024kgu,
    author = "Venugopalan, Gautam and others",
    title = "{Search for new interactions at the micron scale with a vector force sensor}",
    eprint = "2412.13167",
    archivePrefix = "arXiv",
    primaryClass = "hep-ex",
    month = "12",
    year = "2024"
}

@article{Cong:2024qly,
    author = "Cong, Lei and others",
    title = "{Spin-dependent exotic interactions}",
    eprint = "2408.15691",
    archivePrefix = "arXiv",
    primaryClass = "hep-ph",
    month = "8",
    year = "2024"
}

@article{Cyncynates:2024ufu,
    author = "Cyncynates, David and Simon, Olivier",
    title = "{Scalar relics from the hot Big Bang}",
    eprint = "2410.22409",
    archivePrefix = "arXiv",
    primaryClass = "hep-ph",
    month = "10",
    year = "2024"
}

@article{Mostepanenko:2020lqe,
    author = "Mostepanenko, Vladimir M. and Klimchitskaya, Galina L.",
    editor = "Mostepanenko, Vladimir M. and Starobinsky, Alexei A. and Velichko, Elena N.",
    title = "{The State of the Art in Constraining Axion-to-Nucleon Coupling and Non-Newtonian Gravity from Laboratory Experiments}",
    eprint = "2009.04517",
    archivePrefix = "arXiv",
    primaryClass = "hep-ph",
    doi = "10.3390/universe6090147",
    journal = "Universe",
    volume = "6",
    number = "9",
    pages = "147",
    year = "2020"
}

@article{Klimchitskaya:2013rwd,
    author = "Klimchitskaya, G. L. and Mohideen, U. and Mostepanenko, V. M.",
    title = "{Constraints on corrections to Newtonian gravity from two recent measurements of the Casimir interaction between metallic surfaces}",
    eprint = "1306.4979",
    archivePrefix = "arXiv",
    primaryClass = "gr-qc",
    doi = "10.1103/PhysRevD.87.125031",
    journal = "Phys. Rev. D",
    volume = "87",
    number = "12",
    pages = "125031",
    year = "2013"
}

@misc{AxionLimits,
  author       = {Ciaran O'Hare},
  title        = {cajohare/AxionLimits: AxionLimits},
  month        = jul,
  year         = 2020,
  publisher    = {Zenodo},
  version      = {v1.0},
  doi          = {10.5281/zenodo.3932430},
  howpublished = {\url{https://cajohare.github.io/AxionLimits/}}
}

@article{OHare:2020wah,
    author = "O'Hare, Ciaran A. J. and Vitagliano, Edoardo",
    title = "{Cornering the axion with $CP$-violating interactions}",
    eprint = "2010.03889",
    archivePrefix = "arXiv",
    primaryClass = "hep-ph",
    reportNumber = "CPPC-2020-16",
    doi = "10.1103/PhysRevD.102.115026",
    journal = "Phys. Rev. D",
    volume = "102",
    number = "11",
    pages = "115026",
    year = "2020"
}

@article{Banishev:2014jka,
    author = "Banishev, A. A. and Wagner, J. and Emig, T. and Zandi, R. and Mohideen, U.",
    title = "{Experimental and theoretical investigation of the angular dependence of the Casimir force between sinusoidally corrugated surfaces}",
    eprint = "1402.2716",
    archivePrefix = "arXiv",
    primaryClass = "quant-ph",
    doi = "10.1103/PhysRevB.89.235436",
    journal = "Phys. Rev. B",
    volume = "89",
    number = "23",
    pages = "235436",
    year = "2014"
}

@article{Banishev:2012kkb,
    author = "Banishev, A. A. and Wagner, J. and Emig, T. and Zandi, R. and Mohideen, U.",
    title = "{Demonstration of Angle Dependent Casimir Force Between Corrugations}",
    eprint = "1212.6271",
    archivePrefix = "arXiv",
    primaryClass = "quant-ph",
    doi = "10.1103/PhysRevLett.110.250403",
    journal = "Phys. Rev. Lett.",
    volume = "110",
    number = "25",
    pages = "250403",
    year = "2013"
}

@article{Bezerra:2010pq,
    author = "Bezerra, V. B. and Klimchitskaya, G. L. and Mostepanenko, V. M. and Romero, C.",
    title = "{Advance and prospects in constraining the Yukawa-type corrections to Newtonian gravity from the Casimir effect}",
    eprint = "1002.2141",
    archivePrefix = "arXiv",
    primaryClass = "hep-th",
    doi = "10.1103/PhysRevD.81.055003",
    journal = "Phys. Rev. D",
    volume = "81",
    pages = "055003",
    year = "2010"
}

@article{Chiu:2009fqu,
    author = "Chiu, H. -C. and Klimchitskaya, G. L. and Marachevsky, V. N. and Mostepanenko, V. M. and Mohideen, U.",
    title = "{Demonstration of the asymmetric lateral Casimir force between corrugated surfaces in the nonadditive regime}",
    eprint = "0909.2161",
    archivePrefix = "arXiv",
    primaryClass = "quant-ph",
    doi = "10.1103/PhysRevB.80.121402",
    journal = "Phys. Rev. B",
    volume = "80",
    number = "12",
    pages = "121402",
    year = "2009"
}

@article{Kamiya:2015eva,
    author = "Kamiya, Y. and Itagami, K. and Tani, M. and Kim, G. N. and Komamiya, S.",
    title = "{Constraints on New Gravitylike Forces in the Nanometer Range}",
    eprint = "1504.02181",
    archivePrefix = "arXiv",
    primaryClass = "hep-ex",
    doi = "10.1103/PhysRevLett.114.161101",
    journal = "Phys. Rev. Lett.",
    volume = "114",
    pages = "161101",
    year = "2015"
}

@article{Heacock:2021btd,
    author = "Heacock, Benjamin and others",
    title = {{Pendell\"osung interferometry probes the neutron charge radius, lattice dynamics, and fifth forces}},
    eprint = "2103.05428",
    archivePrefix = "arXiv",
    primaryClass = "nucl-ex",
    doi = "10.1126/science.abc2794",
    journal = "Science",
    volume = "373",
    number = "6560",
    pages = "abc2794",
    year = "2021"
}

@article{Haddock:2017wav,
    author = "Haddock, Christopher C. and others",
    title = "{Search for deviations from the inverse square law of gravity at nm range using a pulsed neutron beam}",
    eprint = "1712.02984",
    archivePrefix = "arXiv",
    primaryClass = "nucl-ex",
    doi = "10.1103/PhysRevD.97.062002",
    journal = "Phys. Rev. D",
    volume = "97",
    number = "6",
    pages = "062002",
    year = "2018"
}

@article{Klimchitskaya:2023cgy,
    author = "Klimchitskaya, Galina L. and Mostepanenko, Vladimir M.",
    title = "{How to Strengthen Constraints on Non-Newtonian Gravity from Measuring the Lateral Casimir Force}",
    eprint = "2305.07884",
    archivePrefix = "arXiv",
    primaryClass = "quant-ph",
    doi = "10.3390/universe9010034",
    journal = "Universe",
    volume = "9",
    number = "1",
    pages = "34",
    year = "2023"
}

@article{Grifols:1986fc,
    author = "Grifols, J. A. and Masso, E.",
    title = "{Constraints on Finite Range Baryonic and Leptonic Forces From Stellar Evolution}",
    reportNumber = "UAB-FT-142",
    doi = "10.1016/0370-2693(86)90509-5",
    journal = "Phys. Lett. B",
    volume = "173",
    pages = "237--240",
    year = "1986"
}

@article{Ishizuka:1989ts,
    author = "Ishizuka, N. and Yoshimura, M.",
    title = "{Axion and Dilaton Emissivity From Nascent Neutron Stars}",
    reportNumber = "TU-89-349",
    doi = "10.1143/PTP.84.233",
    journal = "Prog. Theor. Phys.",
    volume = "84",
    pages = "233--250",
    year = "1990"
}

@article{Grifols:1988fv,
    author = "Grifols, J. A. and Masso, E. and Peris, S.",
    title = "{Energy Loss From the Sun and {RED} Giants: Bounds on Short Range Baryonic and Leptonic Forces}",
    reportNumber = "UAB-FT-161-REV, UAB-FT-161",
    doi = "10.1142/S0217732389000381",
    journal = "Mod. Phys. Lett. A",
    volume = "4",
    pages = "311",
    year = "1989"
}

@article{Fiorillo:2023frv,
    author = "Fiorillo, Damiano F. G. and Heinlein, Malte and Janka, Hans-Thomas and Raffelt, Georg and Vitagliano, Edoardo and Bollig, Robert",
    title = "{Supernova simulations confront SN 1987A neutrinos}",
    eprint = "2308.01403",
    archivePrefix = "arXiv",
    primaryClass = "astro-ph.HE",
    doi = "10.1103/PhysRevD.108.083040",
    journal = "Phys. Rev. D",
    volume = "108",
    number = "8",
    pages = "083040",
    year = "2023"
}

@article{Flacke:2016szy,
    author = "Flacke, Thomas and Frugiuele, Claudia and Fuchs, Elina and Gupta, Rick S. and Perez, Gilad",
    title = "{Phenomenology of relaxion-Higgs mixing}",
    eprint = "1610.02025",
    archivePrefix = "arXiv",
    primaryClass = "hep-ph",
    reportNumber = "CTPU-16-25",
    doi = "10.1007/JHEP06(2017)050",
    journal = "JHEP",
    volume = "06",
    pages = "050",
    year = "2017"
}

@article{Chen:2014oda,
    author = "Chen, Y. -J. and Tham, W. K. and Krause, D. E. and Lopez, D. and Fischbach, Ephraim and Decca, R. S.",
    title = "{Stronger Limits on Hypothetical Yukawa Interactions in the 30\textendash{}8000 nm Range}",
    eprint = "1410.7267",
    archivePrefix = "arXiv",
    primaryClass = "hep-ex",
    doi = "10.1103/PhysRevLett.116.221102",
    journal = "Phys. Rev. Lett.",
    volume = "116",
    number = "22",
    pages = "221102",
    year = "2016"
}

@article{Geraci:2008hb,
    author = "Geraci, Andrew A. and Smullin, Sylvia J. and Weld, David M. and Chiaverini, John and Kapitulnik, Aharon",
    title = "{Improved constraints on non-Newtonian forces at 10 microns}",
    eprint = "0802.2350",
    archivePrefix = "arXiv",
    primaryClass = "hep-ex",
    doi = "10.1103/PhysRevD.78.022002",
    journal = "Phys. Rev. D",
    volume = "78",
    pages = "022002",
    year = "2008"
}

@article{Ke:2021jtj,
    author = "Ke, Jun and Luo, Jie and Shao, Cheng-Gang and Tan, Yu-Jie and Tan, Wen-Hai and Yang, Shan-Qing",
    title = "{Combined Test of the Gravitational Inverse-Square Law at the Centimeter Range}",
    doi = "10.1103/PhysRevLett.126.211101",
    journal = "Phys. Rev. Lett.",
    volume = "126",
    number = "21",
    pages = "211101",
    year = "2021"
}

@article{Kapner:2006si,
    author = "Kapner, D. J. and Cook, T. S. and Adelberger, E. G. and Gundlach, J. H. and Heckel, Blayne R. and Hoyle, C. D. and Swanson, H. E.",
    title = "{Tests of the gravitational inverse-square law below the dark-energy length scale}",
    eprint = "hep-ph/0611184",
    archivePrefix = "arXiv",
    doi = "10.1103/PhysRevLett.98.021101",
    journal = "Phys. Rev. Lett.",
    volume = "98",
    pages = "021101",
    year = "2007"
}

@article{Lee:2020zjt,
    author = "Lee, J. G. and Adelberger, E. G. and Cook, T. S. and Fleischer, S. M. and Heckel, B. R.",
    title = "{New Test of the Gravitational $1/r^2$ Law at Separations down to 52 $\mu$m}",
    eprint = "2002.11761",
    archivePrefix = "arXiv",
    primaryClass = "hep-ex",
    doi = "10.1103/PhysRevLett.124.101101",
    journal = "Phys. Rev. Lett.",
    volume = "124",
    number = "10",
    pages = "101101",
    year = "2020"
}

@article{Wei:2022ggs,
    author = "Wei, Kai and Zhao, Tian and Fang, Xiujie and Xu, Zitong and Liu, Chang and Cao, Qian and Wickenbrock, Arne and Hu, Yanhui and Ji, Wei and Budker, Dmitry",
    title = "{Ultrasensitive Atomic Comagnetometer with Enhanced Nuclear Spin Coherence}",
    eprint = "2210.09027",
    archivePrefix = "arXiv",
    primaryClass = "physics.atom-ph",
    doi = "10.1103/PhysRevLett.130.063201",
    journal = "Phys. Rev. Lett.",
    volume = "130",
    number = "6",
    pages = "063201",
    year = "2023"
}

@article{Feng:2022tsu,
    author = "Feng, Yukun and Ning, Denghui and Zhang, Shaobo and Lu, Zhengtian and Sheng, Dong",
    title = "{Search for Monopole-Dipole Interactions at the Submillimeter Range with a Xe129-Xe131-Rb Comagnetometer}",
    eprint = "2205.13237",
    archivePrefix = "arXiv",
    primaryClass = "physics.atom-ph",
    doi = "10.1103/PhysRevLett.128.231803",
    journal = "Phys. Rev. Lett.",
    volume = "128",
    number = "23",
    pages = "231803",
    year = "2022"
}

@article{Fan:2023hci,
    author = "Fan, Xing and Reig, Mario",
    title = "{New bounds and future prospects for axion force searches at Penning trap experiments}",
    eprint = "2310.18797",
    archivePrefix = "arXiv",
    primaryClass = "hep-ph",
    month = "10",
    year = "2023"
}

@article{Wu:2023ypz,
    author = "Wu, L. Y. and Zhang, K. Y. and Peng, M. and Gong, J. and Yan, H.",
    title = "{New Limits on Exotic Spin-Dependent Interactions at Astronomical Distances}",
    eprint = "2302.09096",
    archivePrefix = "arXiv",
    primaryClass = "hep-ph",
    doi = "10.1103/PhysRevLett.131.091002",
    journal = "Phys. Rev. Lett.",
    volume = "131",
    number = "9",
    pages = "091002",
    year = "2023"
}

@article{Agrawal:2023lmw,
    author = "Agrawal, Prateek and Hutzler, Nicholas R. and Kaplan, David E. and Rajendran, Surjeet and Reig, Mario",
    title = "{Searching for axion forces with spin precession in atoms and molecules}",
    eprint = "2309.10023",
    archivePrefix = "arXiv",
    primaryClass = "hep-ph",
    reportNumber = "FERMILAB-PUB-23-651-SQMS-V",
    doi = "10.1007/JHEP07(2024)133",
    journal = "JHEP",
    volume = "07",
    pages = "133",
    year = "2024"
}

@article{Capozzi:2020cbu,
    author = "Capozzi, Francesco and Raffelt, Georg",
    title = "{Axion and neutrino bounds improved with new calibrations of the tip of the red-giant branch using geometric distance determinations}",
    eprint = "2007.03694",
    archivePrefix = "arXiv",
    primaryClass = "astro-ph.SR",
    reportNumber = "MPP-2020-106",
    doi = "10.1103/PhysRevD.102.083007",
    journal = "Phys. Rev. D",
    volume = "102",
    number = "8",
    pages = "083007",
    year = "2020"
}

@article{Berge:2021yye,
    author = "Berg\'e, Joel and Pernot-Borr\`as, Martin and Uzan, Jean-Philippe and Brax, Philippe and Chhun, Ratana and M\'etris, Gilles and Rodrigues, Manuel and Touboul, Pierre",
    title = "{MICROSCOPE\textquoteright{}s constraint on a short-range fifth force}",
    eprint = "2102.00022",
    archivePrefix = "arXiv",
    primaryClass = "gr-qc",
    doi = "10.1088/1361-6382/abe142",
    journal = "Class. Quant. Grav.",
    volume = "39",
    number = "20",
    pages = "204010",
    year = "2022"
}

@article{Dekens:2022gha,
    author = "Dekens, Wouter and de Vries, Jordy and Shain, Sachin",
    title = "{CP-violating axion interactions in effective field theory}",
    eprint = "2203.11230",
    archivePrefix = "arXiv",
    primaryClass = "hep-ph",
    doi = "10.1007/JHEP07(2022)014",
    journal = "JHEP",
    volume = "07",
    pages = "014",
    year = "2022"
}

@article{MICROSCOPE:2022doy,
    author = "Touboul, Pierre and others",
    collaboration = "MICROSCOPE",
    title = "{MICROSCOPE Mission: Final Results of the Test of the Equivalence Principle}",
    eprint = "2209.15487",
    archivePrefix = "arXiv",
    primaryClass = "gr-qc",
    doi = "10.1103/PhysRevLett.129.121102",
    journal = "Phys. Rev. Lett.",
    volume = "129",
    number = "12",
    pages = "121102",
    year = "2022"
}

@article{Cyncynates:2024bxw,
    author = "Cyncynates, David and Simon, Olivier",
    title = "{Minimal targets for dilaton direct detection}",
    eprint = "2408.16816",
    archivePrefix = "arXiv",
    primaryClass = "hep-ph",
    month = "8",
    year = "2024"
}

@article{DiLuzio:2024ctr,
    author = "Di Luzio, Luca and Gisbert, Hector and Nesti, Fabrizio and S\o{}rensen, Philip",
    title = "{Axion window on new macroscopic forces}",
    eprint = "2407.15928",
    archivePrefix = "arXiv",
    primaryClass = "hep-ph",
    doi = "10.1103/PhysRevD.110.115034",
    journal = "Phys. Rev. D",
    volume = "110",
    number = "11",
    pages = "115034",
    year = "2024"
}

@article{Smith:1999cr,
    author = "Smith, G. L. and Hoyle, C. D. and Gundlach, J. H. and Adelberger, E. G. and Heckel, Blayne R. and Swanson, H. E.",
    title = "{Short range tests of the equivalence principle}",
    eprint = "2405.10982",
    archivePrefix = "arXiv",
    primaryClass = "gr-qc",
    doi = "10.1103/PhysRevD.61.022001",
    journal = "Phys. Rev. D",
    volume = "61",
    pages = "022001",
    year = "2000"
}

@article{Yang:2012zzb,
    author = "Yang, Shan-Qing and Zhan, Bi-Fu and Wang, Qing-Lan and Shao, Cheng-Gang and Tu, Liang-Cheng and Tan, Wen-Hai and Luo, Jun",
    title = "{Test of the Gravitational Inverse Square Law at Millimeter Ranges}",
    doi = "10.1103/PhysRevLett.108.081101",
    journal = "Phys. Rev. Lett.",
    volume = "108",
    pages = "081101",
    year = "2012"
}

@article{Tan:2020vpf,
    author = "Tan, Wen-Hai and others",
    title = "{Improvement for Testing the Gravitational Inverse-Square Law at the Submillimeter Range}",
    doi = "10.1103/PhysRevLett.124.051301",
    journal = "Phys. Rev. Lett.",
    volume = "124",
    number = "5",
    pages = "051301",
    year = "2020"
}

@article{Hoskins:1985tn,
    author = "Hoskins, J. K. and Newman, R. D. and Spero, R. and Schultz, J.",
    title = "{Experimental tests of the gravitational inverse square law for mass separations from 2-cm to 105-cm}",
    doi = "10.1103/PhysRevD.32.3084",
    journal = "Phys. Rev. D",
    volume = "32",
    pages = "3084--3095",
    year = "1985"
}

@article{Hardy:2016kme,
    author = "Hardy, Edward and Lasenby, Robert",
    title = "{Stellar cooling bounds on new light particles: plasma mixing effects}",
    eprint = "1611.05852",
    archivePrefix = "arXiv",
    primaryClass = "hep-ph",
    doi = "10.1007/JHEP02(2017)033",
    journal = "JHEP",
    volume = "02",
    pages = "033",
    year = "2017"
}

@article{Berge:2017ovy,
    author = "Berg\'e, Joel and Brax, Philippe and M\'etris, Gilles and Pernot-Borr\`as, Martin and Touboul, Pierre and Uzan, Jean-Philippe",
    title = "{MICROSCOPE Mission: First Constraints on the Violation of the Weak Equivalence Principle by a Light Scalar Dilaton}",
    eprint = "1712.00483",
    archivePrefix = "arXiv",
    primaryClass = "gr-qc",
    doi = "10.1103/PhysRevLett.120.141101",
    journal = "Phys. Rev. Lett.",
    volume = "120",
    number = "14",
    pages = "141101",
    year = "2018"
}

@article{Damour:1994zq,
    author = "Damour, T. and Polyakov, Alexander M.",
    title = "{The String dilaton and a least coupling principle}",
    eprint = "hep-th/9401069",
    archivePrefix = "arXiv",
    doi = "10.1016/0550-3213(94)90143-0",
    journal = "Nucl. Phys. B",
    volume = "423",
    pages = "532--558",
    year = "1994"
}

@article{Taylor:1988nw,
    author = "Taylor, T. R. and Veneziano, G.",
    title = "{Dilaton Couplings at Large Distances}",
    reportNumber = "CERN-TH-5116-88, FERMILAB-PUB-88-089-T",
    doi = "10.1016/0370-2693(88)91290-7",
    journal = "Phys. Lett. B",
    volume = "213",
    pages = "450--454",
    year = "1988"
}

@article{Hui:2016ltb,
    author = "Hui, Lam and Ostriker, Jeremiah P. and Tremaine, Scott and Witten, Edward",
    title = "{Ultralight scalars as cosmological dark matter}",
    eprint = "1610.08297",
    archivePrefix = "arXiv",
    primaryClass = "astro-ph.CO",
    doi = "10.1103/PhysRevD.95.043541",
    journal = "Phys. Rev. D",
    volume = "95",
    number = "4",
    pages = "043541",
    year = "2017"
}

@book{Raffelt:1996wa,
  author    = {Georg G. Raffelt},
  title     = {Stars as Laboratories for Fundamental Physics: The Astrophysics of Neutrinos, Axions, and Other Weakly Interacting Particles},
  publisher = {University of Chicago Press},
  address   = {Chicago},
  year      = {1996},
  isbn      = {0-226-70272-3},
  note      = {Theoretical Astrophysics; 686~pp., 188 line drawings, 34 tables}
}

@article{Knapen:2017xzo,
    author = "Knapen, Simon and Lin, Tongyan and Zurek, Kathryn M.",
    title = "{Light Dark Matter: Models and Constraints}",
    eprint = "1709.07882",
    archivePrefix = "arXiv",
    primaryClass = "hep-ph",
    doi = "10.1103/PhysRevD.96.115021",
    journal = "Phys. Rev. D",
    volume = "96",
    number = "11",
    pages = "115021",
    year = "2017"
}

@article{Arkani-Hamed:1999lsd,
    author = "Arkani-Hamed, Nima and Hall, Lawrence J. and Tucker-Smith, David and Weiner, Neal",
    title = "{Solving the hierarchy problem with exponentially large dimensions}",
    eprint = "hep-ph/9912453",
    archivePrefix = "arXiv",
    reportNumber = "LBNL-44717, UCB-PTH-99-55, LBL-44717",
    doi = "10.1103/PhysRevD.62.105002",
    journal = "Phys. Rev. D",
    volume = "62",
    pages = "105002",
    year = "2000"
}

@article{Will:2014kxa,
    author = "Will, Clifford M.",
    title = "{The Confrontation between General Relativity and Experiment}",
    eprint = "1403.7377",
    archivePrefix = "arXiv",
    primaryClass = "gr-qc",
    doi = "10.12942/lrr-2014-4",
    journal = "Living Rev. Rel.",
    volume = "17",
    pages = "4",
    year = "2014"
}

@article{Baruch:2024frj,
    author = "Baruch, Chaja and Changala, P. Bryan and Shagam, Yuval and Soreq, Yotam",
    title = "{Constraining CP Violating Nucleon-Nucleon Long-Range Interactions in Diatomic eEDM Searches}",
    eprint = "2402.07504",
    archivePrefix = "arXiv",
    primaryClass = "hep-ph",
    doi = "10.1103/PhysRevLett.133.113202",
    journal = "Phys. Rev. Lett.",
    volume = "133",
    number = "11",
    pages = "113202",
    year = "2024"
}

@article{DiLuzio:2023lmd,
    author = "Di Luzio, Luca and Gisbert, Hector and Levati, Gabriele and Paradisi, Paride and S\o{}rensen, Philip",
    title = "{CP-Violating Axions: A Theory Review}",
    eprint = "2312.17310",
    archivePrefix = "arXiv",
    primaryClass = "hep-ph",
    month = "12",
    year = "2023"
}

@article{Lee:2018vaq,
    author = "Lee, Junyi and Almasi, Attaallah and Romalis, Michael",
    title = "{Improved Limits on Spin-Mass Interactions}",
    eprint = "1801.02757",
    archivePrefix = "arXiv",
    primaryClass = "hep-ex",
    doi = "10.1103/PhysRevLett.120.161801",
    journal = "Phys. Rev. Lett.",
    volume = "120",
    number = "16",
    pages = "161801",
    year = "2018"
}

@article{Venema:1992zz,
    author = "Venema, B.J. and Majumder, P.K. and Lamoreaux, S.K. and Heckel, B.R. and Fortson, E.N.",
    title = "{Search for a coupling of the Earth's gravitational field to nuclear spins in atomic mercury}",
    doi = "10.1103/PhysRevLett.68.135",
    journal = "Phys. Rev. Lett.",
    volume = "68",
    pages = "135--138",
    year = "1992"
}

@misc{supplementalmaterial,
  note = {See Supplemental Material for a discussion concerning production of scalars in cold neutron stars, additional information on neutron star observations, and an update of current bounds on CP-violating axions. It includes Refs.~\cite{Ericson:1988wr, Thorsett:2003xy, Abel:2020pzs, Cong:2024qly, Jain:2005nh, Masso:2005ym, Jaeckel:2006xm, Masso:2006gc, Budnik:2020nwz, DeRocco:2020xdt}.},
  howpublished = {\url{https://}} 
}

@article{Pospelov:2005pr,
      author         = "Pospelov, Maxim and Ritz, Adam",
      title          = "{Electric dipole moments as probes of new physics}",
      journal        = "Annals Phys.",
      volume         = "318",
      year           = "2005",
      pages          = "119-169",
      doi            = "10.1016/j.aop.2005.04.002",
      eprint         = "hep-ph/0504231",
      archivePrefix  = "arXiv",
      primaryClass   = "hep-ph",
      SLACcitation   = "%%CITATION = HEP-PH/0504231;%%"
}

@article{Masso:2006gc,
    author = "Masso, Eduard and Redondo, Javier",
    title = "{Compatibility of CAST search with axion-like interpretation of PVLAS results}",
    eprint = "hep-ph/0606163",
    archivePrefix = "arXiv",
    reportNumber = "UAB-FT-605",
    doi = "10.1103/PhysRevLett.97.151802",
    journal = "Phys. Rev. Lett.",
    volume = "97",
    pages = "151802",
    year = "2006"
}

@article{Budnik:2020nwz,
    author = "Budnik, Ranny and Kim, Hyungjin and Matsedonskyi, Oleksii and Perez, Gilad and Soreq, Yotam",
    title = "{Probing the relaxed relaxion and Higgs-portal with S1 \& S2}",
    eprint = "2006.14568",
    archivePrefix = "arXiv",
    primaryClass = "hep-ph",
    month = "6",
    year = "2020"
}

@article{Bloch:2020uzh,
    author = "Bloch, Itay M. and Caputo, Andrea and Essig, Rouven and Redigolo, Diego and Sholapurkar, Mukul and Volansky, Tomer",
    title = "{Exploring new physics with O(keV) electron recoils in direct detection experiments}",
    eprint = "2006.14521",
    archivePrefix = "arXiv",
    primaryClass = "hep-ph",
    doi = "10.1007/JHEP01(2021)178",
    journal = "JHEP",
    volume = "01",
    pages = "178",
    year = "2021"
}

@article{DeRocco:2020xdt,
    author = "DeRocco, William and Graham, Peter W. and Rajendran, Surjeet",
    title = "{Exploring the robustness of stellar cooling constraints on light particles}",
    eprint = "2006.15112",
    archivePrefix = "arXiv",
    primaryClass = "hep-ph",
    doi = "10.1103/PhysRevD.102.075015",
    journal = "Phys. Rev. D",
    volume = "102",
    number = "7",
    pages = "075015",
    year = "2020"
}

@article{Jaeckel:2006xm,
    author = "Jaeckel, Joerg and Masso, Eduard and Redondo, Javier and Ringwald, Andreas and Takahashi, Fuminobu",
    title = "{The Need for purely laboratory-based axion-like particle searches}",
    eprint = "hep-ph/0610203",
    archivePrefix = "arXiv",
    reportNumber = "DCPT-06-136, DESY-06-188, IPPP-06-68, UAB-FT-612",
    doi = "10.1103/PhysRevD.75.013004",
    journal = "Phys. Rev. D",
    volume = "75",
    pages = "013004",
    year = "2007"
}

@article{Masso:2005ym,
    author = "Masso, Eduard and Redondo, Javier",
    title = "{Evading astrophysical constraints on axion-like particles}",
    eprint = "hep-ph/0504202",
    archivePrefix = "arXiv",
    reportNumber = "UAB-FT-579",
    doi = "10.1088/1475-7516/2005/09/015",
    journal = "JCAP",
    volume = "09",
    pages = "015",
    year = "2005"
}

@article{Jain:2005nh,
    author = "Jain, Pankaj and Mandal, Subhayan",
    title = "{Evading the astrophysical limits on light pseudoscalars}",
    eprint = "astro-ph/0512155",
    archivePrefix = "arXiv",
    doi = "10.1142/S0218271806009558",
    journal = "Int. J. Mod. Phys. D",
    volume = "15",
    pages = "2095--2104",
    year = "2006"
}

@article{Raffelt:2012sp,
    author = "Raffelt, Georg",
    title = "{Limits on a CP-violating scalar axion-nucleon interaction}",
    eprint = "1205.1776",
    archivePrefix = "arXiv",
    primaryClass = "hep-ph",
    reportNumber = "MPP-2012-74",
    doi = "10.1103/PhysRevD.86.015001",
    journal = "Phys. Rev. D",
    volume = "86",
    pages = "015001",
    year = "2012"
}

@article{Bertolini:2020hjc,
    author = "Bertolini, Stefano and Di Luzio, Luca and Nesti, Fabrizio",
    title = "{Axion-mediated forces and CP violation in left-right models}",
    eprint = "2006.12508",
    archivePrefix = "arXiv",
    primaryClass = "hep-ph",
    reportNumber = "DESY-20-103",
    month = "6",
    year = "2020"
}

@article{Bigazzi:2019hav,
    author = {Bigazzi, Francesco and Cotrone, Aldo L. and J\"arvinen, Matti and Kiritsis, Elias},
    title = "{Non-derivative Axionic Couplings to Nucleons at large and small N}",
    eprint = "1906.12132",
    archivePrefix = "arXiv",
    primaryClass = "hep-ph",
    reportNumber = "CCTP-2019-6; ITCP-IPP 2019/6, CCTP-2019-6, ITCP-IPP 2019/6",
    doi = "10.1007/JHEP01(2020)100",
    journal = "JHEP",
    volume = "01",
    pages = "100",
    year = "2020"
}

@article{Pospelov:1997uv,
      author         = "Pospelov, M.",
      title          = "{CP odd interaction of axion with matter}",
      journal        = "Phys. Rev. D",
      volume         = "58",
      year           = "1998",
      pages          = "097703",
      doi            = "10.1103/PhysRevD.58.097703",
      eprint         = "hep-ph/9707431",
      archivePrefix  = "arXiv",
      primaryClass   = "hep-ph",
      reportNumber   = "UQAM-PHE-97-04",
      SLACcitation   = "%%CITATION = HEP-PH/9707431;%%"
}

@article{Georgi:1986kr,
      author         = "Georgi, Howard and Randall, Lisa",
      title          = "{Flavor Conserving CP Violation in Invisible Axion
                        Models}",
      journal        = "Nucl. Phys. B",
      volume         = "276",
      year           = "1986",
      pages          = "241-252",
      doi            = "10.1016/0550-3213(86)90022-2",
      reportNumber   = "HUTP-86/A025",
      SLACcitation   = "%%CITATION = NUPHA,B276,241;%%"
}

@article{Moody:1984ba,
      author         = "Moody, J. E. and Wilczek, Frank",
      title          = "{New Macroscopic Forces?}",
      journal        = "Phys. Rev. D",
      volume         = "30",
      year           = "1984",
      pages          = "130",
      doi            = "10.1103/PhysRevD.30.130",
      reportNumber   = "NSF-ITP-83-177",
      SLACcitation   = "%%CITATION = PHRVA,D30,130;%%"
}

@article{Geraci:2017bmq,
    author = "Geraci, A.A. and others",
    editor = "Carosi, Gianpaolo and Rybka, Gray and van Bibber, Karl",
    collaboration = "ARIADNE",
    title = "{Progress on the ARIADNE axion experiment}",
    eprint = "1710.05413",
    archivePrefix = "arXiv",
    primaryClass = "astro-ph.IM",
    doi = "10.1007/978-3-319-92726-8_18",
    journal = "Springer Proc. Phys.",
    volume = "211",
    pages = "151--161",
    year = "2018"
}

@article{Arvanitaki:2014dfa,
    author = "Arvanitaki, Asimina and Geraci, Andrew A.",
    title = "{Resonantly Detecting Axion-Mediated Forces with Nuclear Magnetic Resonance}",
    eprint = "1403.1290",
    archivePrefix = "arXiv",
    primaryClass = "hep-ph",
    doi = "10.1103/PhysRevLett.113.161801",
    journal = "Phys. Rev. Lett.",
    volume = "113",
    number = "16",
    pages = "161801",
    year = "2014"
}

@article{Agrawal:2022wjm,
    author = "Agrawal, Prateek and Kaplan, David E. and Kim, On and Rajendran, Surjeet and Reig, Mario",
    title = "{Searching for axion forces with precision precession in storage rings}",
    eprint = "2210.17547",
    archivePrefix = "arXiv",
    primaryClass = "hep-ph",
    reportNumber = "FERMILAB-PUB-22-847-V",
    doi = "10.1103/PhysRevD.108.015017",
    journal = "Phys. Rev. D",
    volume = "108",
    number = "1",
    pages = "015017",
    year = "2023"
}

@article{Tullney:2013wqa,
    author = "Tullney, K. and others",
    title = "{Constraints on Spin-Dependent Short-Range Interaction between Nucleons}",
    eprint = "1303.6612",
    archivePrefix = "arXiv",
    primaryClass = "hep-ex",
    doi = "10.1103/PhysRevLett.111.100801",
    journal = "Phys. Rev. Lett.",
    volume = "111",
    pages = "100801",
    year = "2013"
}

@article{Crescini:2016lwj,
    author = "Crescini, Nicol\`o and Braggio, Caterina and Carugno, Giovanni and Falferi, Paolo and Ortolan, Antonello and Ruoso, Giuseppe",
    title = "{The QUAX-g$_p$ g$_s$ experiment to search for monopole-dipole Axion interaction}",
    eprint = "1606.04751",
    archivePrefix = "arXiv",
    primaryClass = "physics.ins-det",
    doi = "10.1016/j.nima.2016.10.050",
    journal = "Nucl. Instrum. Meth. A",
    volume = "842",
    pages = "109--113",
    year = "2017"
}

@article{Wineland:1991zz,
      author         = "Wineland, D. J. and Bollinger, J. J. and Heinzen, D. J.
                        and Itano, W. M. and Raizen, M. G.",
      title          = "{Search for anomalous spin-dependent forces using
                        stored-ion spectroscopy}",
      journal        = "Phys. Rev. Lett.",
      volume         = "67",
      year           = "1991",
      pages          = "1735-1738",
      doi            = "10.1103/PhysRevLett.67.1735",
      SLACcitation   = "%%CITATION = PRLTA,67,1735;%%"
}

@article{Okawa:2021fto,
    author = "Okawa, Shohei and Pospelov, Maxim and Ritz, Adam",
    title = "{Long-range axion forces and hadronic CP violation}",
    eprint = "2111.08040",
    archivePrefix = "arXiv",
    primaryClass = "hep-ph",
    doi = "10.1103/PhysRevD.105.075003",
    journal = "Phys. Rev. D",
    volume = "105",
    number = "7",
    pages = "075003",
    year = "2022"
}

@article{Crescini:2017uxs,
    author = "Crescini, N. and Braggio, C. and Carugno, G. and Falferi, P. and Ortolan, A. and Ruoso, G.",
    title = "{Improved constraints on monopole-dipole interaction mediated by pseudo-scalar bosons}",
    eprint = "1705.06044",
    archivePrefix = "arXiv",
    primaryClass = "hep-ex",
    doi = "10.1016/j.physletb.2017.09.019",
    journal = "Phys. Lett. B",
    volume = "773",
    pages = "677--680",
    year = "2017"
}

@article{Heckel:2008hw,
    author = "Heckel, Blayne R. and Adelberger, E.G. and Cramer, C.E. and Cook, T.S. and Schlamminger, Stephan and Schmidt, U.",
    title = "{Preferred-Frame and CP-Violation Tests with Polarized Electrons}",
    eprint = "0808.2673",
    archivePrefix = "arXiv",
    primaryClass = "hep-ex",
    doi = "10.1103/PhysRevD.78.092006",
    journal = "Phys. Rev. D",
    volume = "78",
    pages = "092006",
    year = "2008"
}

@article{Tan:2016vwu,
    author = "Tan, Wen-Hai and Yang, Shan-Qing and Shao, Cheng-Gang and Li, Jia and Du, An-Bin and Zhan, Bi-Fu and Wang, Qing-Lan and Luo, Peng-Shun and Tu, Liang-Cheng and Luo, Jun",
    title = "{New Test of the Gravitational Inverse-Square Law at the Submillimeter Range with Dual Modulation and Compensation}",
    doi = "10.1103/PhysRevLett.116.131101",
    journal = "Phys. Rev. Lett.",
    volume = "116",
    number = "13",
    pages = "131101",
    year = "2016"
}

@article{Tu:2007zz,
    author = "Tu, Liang-Cheng and Guan, Sheng-Guo and Luo, Jun and Shao, Cheng-Gang and Liu, Lin-Xia",
    title = "{Null Test of Newtonian Inverse-Square Law at Submillimeter Range with a Dual-Modulation Torsion Pendulum}",
    doi = "10.1103/PhysRevLett.98.201101",
    journal = "Phys. Rev. Lett.",
    volume = "98",
    pages = "201101",
    year = "2007"
}

@article{Plakkot:2023pui,
    author = "Plakkot, V. and Dekens, W. and de Vries, J. and Shain, S.",
    title = "{CP-violating axion interactions II: axions as dark matter}",
    eprint = "2306.07065",
    archivePrefix = "arXiv",
    primaryClass = "hep-ph",
    doi = "10.1007/JHEP11(2023)012",
    journal = "JHEP",
    volume = "11",
    pages = "012",
    year = "2023"
}

@article{DiLuzio:2023cuk,
    author = "Di Luzio, Luca and Levati, Gabriele and Paradisi, Paride",
    title = "{The chiral Lagrangian of CP-violating axion-like particles}",
    eprint = "2311.12158",
    archivePrefix = "arXiv",
    primaryClass = "hep-ph",
    doi = "10.1007/JHEP02(2024)020",
    journal = "JHEP",
    volume = "02",
    pages = "020",
    year = "2024"
}

@article{Baruch:2024fbh,
    author = "Baruch, Chaja and Changala, P. Bryan and Shagam, Yuval and Soreq, Yotam",
    title = "{Constraining P and T violating forces with chiral molecules}",
    eprint = "2406.02281",
    archivePrefix = "arXiv",
    primaryClass = "hep-ph",
    doi = "10.1103/PhysRevResearch.6.043115",
    journal = "Phys. Rev. Res.",
    volume = "6",
    number = "4",
    pages = "043115",
    year = "2024"
}

@article{Stadnik:2017hpa,
    author = "Stadnik, Y.V. and Dzuba, V.A. and Flambaum, V.V.",
    title = "{Improved Limits on Axionlike-Particle-Mediated P , T -Violating Interactions between Electrons and Nucleons from Electric Dipole Moments of Atoms and Molecules}",
    eprint = "1708.00486",
    archivePrefix = "arXiv",
    primaryClass = "physics.atom-ph",
    doi = "10.1103/PhysRevLett.120.013202",
    journal = "Phys. Rev. Lett.",
    volume = "120",
    number = "1",
    pages = "013202",
    year = "2018"
}

@article{Gerard:2012ud,
    author = "G\'erard, Jean-Marc and Mertens, Philippe",
    title = "{Weakly-induced strong CP-violation}",
    eprint = "1206.0914",
    archivePrefix = "arXiv",
    primaryClass = "hep-ph",
    reportNumber = "CP3-12-29",
    doi = "10.1016/j.physletb.2012.08.036",
    journal = "Phys. Lett. B",
    volume = "716",
    pages = "316--321",
    year = "2012"
}

@article{Khriplovich:1985jr,
    author = "Khriplovich, I. B.",
    title = "{Quark Electric Dipole Moment and Induced $\theta$ Term in the {Kobayashi-Maskawa} Model}",
    reportNumber = "IYF-86-25",
    doi = "10.1016/0370-2693(86)90245-5",
    journal = "Phys. Lett. B",
    volume = "173",
    pages = "193--196",
    year = "1986"
}

@article{Ellis:1978hq,
    author = "Ellis, John R. and Gaillard, Mary K.",
    title = "{Strong and Weak CP Violation}",
    reportNumber = "FERMILAB-PUB-78-066-T",
    doi = "10.1016/0550-3213(79)90297-9",
    journal = "Nucl. Phys. B",
    volume = "150",
    pages = "141--162",
    year = "1979"
}

@article{Abel:2020pzs,
    author = "Abel, C. and others",
    title = "{Measurement of the Permanent Electric Dipole Moment of the Neutron}",
    eprint = "2001.11966",
    archivePrefix = "arXiv",
    primaryClass = "hep-ex",
    doi = "10.1103/PhysRevLett.124.081803",
    journal = "Phys. Rev. Lett.",
    volume = "124",
    number = "8",
    pages = "081803",
    year = "2020"
}

@article{Adelberger:2003zx,
    author = "Adelberger, E. G. and Heckel, Blayne R. and Nelson, A. E.",
    title = "{Tests of the gravitational inverse square law}",
    eprint = "hep-ph/0307284",
    archivePrefix = "arXiv",
    doi = "10.1146/annurev.nucl.53.041002.110503",
    journal = "Ann. Rev. Nucl. Part. Sci.",
    volume = "53",
    pages = "77--121",
    year = "2003"
}

@article{Tino:2020nla,
    author = "Tino, G. M. and Cacciapuoti, L. and Capozziello, S. and Lambiase, G. and Sorrentino, F.",
    title = "{Precision Gravity Tests and the Einstein Equivalence Principle}",
    eprint = "2002.02907",
    archivePrefix = "arXiv",
    primaryClass = "gr-qc",
    doi = "10.1016/j.ppnp.2020.103772",
    journal = "Prog. Part. Nucl. Phys.",
    volume = "112",
    pages = "103772",
    year = "2020"
}

@article{Krnjaic:2015mbs,
    author = "Krnjaic, Gordan",
    title = "{Probing Light Thermal Dark-Matter With a Higgs Portal Mediator}",
    eprint = "1512.04119",
    archivePrefix = "arXiv",
    primaryClass = "hep-ph",
    reportNumber = "FERMILAB-PUB-15-550-A",
    doi = "10.1103/PhysRevD.94.073009",
    journal = "Phys. Rev. D",
    volume = "94",
    number = "7",
    pages = "073009",
    year = "2016"
}

@article{Potekhin:2020ttj,
    author = "Potekhin, A. Y. and Zyuzin, D. A. and Yakovlev, D. G. and Beznogov, M. V. and Shibanov, Yu. A.",
    title = "{Thermal luminosities of cooling neutron stars}",
    eprint = "2006.15004",
    archivePrefix = "arXiv",
    primaryClass = "astro-ph.HE",
    doi = "10.1093/mnras/staa1871",
    journal = "Mon. Not. Roy. Astron. Soc.",
    volume = "496",
    number = "4",
    pages = "5052--5071",
    year = "2020"
}

@article{Bottaro:2023gep,
    author = "Bottaro, Salvatore and Caputo, Andrea and Raffelt, Georg and Vitagliano, Edoardo",
    title = "{Stellar limits on scalars from electron-nucleus bremsstrahlung}",
    eprint = "2303.00778",
    archivePrefix = "arXiv",
    primaryClass = "hep-ph",
    reportNumber = "CERN-TH-2023-035",
    doi = "10.1088/1475-7516/2023/07/071",
    journal = "JCAP",
    volume = "07",
    pages = "071",
    year = "2023"
}

@article{Ferreira:2022xlw,
    author = {Ferreira, Ricardo Z. and Marsh, M. C. David and M\"uller, Eike},
    title = "{Strong supernovae bounds on ALPs from quantum loops}",
    eprint = "2205.07896",
    archivePrefix = "arXiv",
    primaryClass = "hep-ph",
    doi = "10.1088/1475-7516/2022/11/057",
    journal = "JCAP",
    volume = "11",
    pages = "057",
    year = "2022"
}

@article{Suzuki:2021ium,
    author = "Suzuki, Hiromasa and Bamba, Aya and Shibata, Shinpei",
    title = "{Quantitative Age Estimation of Supernova Remnants and Associated Pulsars}",
    eprint = "2104.10052",
    archivePrefix = "arXiv",
    primaryClass = "astro-ph.HE",
    doi = "10.3847/1538-4357/abfb02",
    journal = "Astrophys. J.",
    volume = "914",
    number = "2",
    pages = "103",
    year = "2021"
}

@article{Schwenk:2003pj,
    author = "Schwenk, Achim and Jaikumar, Prashanth and Gale, Charles",
    title = "{Neutrino bremsstrahlung in neutron matter from effective nuclear interactions}",
    eprint = "nucl-th/0309072",
    archivePrefix = "arXiv",
    doi = "10.1016/j.physletb.2004.01.036",
    journal = "Phys. Lett. B",
    volume = "584",
    pages = "241--250",
    year = "2004"
}

@article{Schwenk:2003bc,
    author = "Schwenk, Achim and Friman, Bengt",
    title = "{Polarization contributions to the spin dependence of the effective interaction in neutron matter}",
    eprint = "nucl-th/0307089",
    archivePrefix = "arXiv",
    doi = "10.1103/PhysRevLett.92.082501",
    journal = "Phys. Rev. Lett.",
    volume = "92",
    pages = "082501",
    year = "2004"
}

@article{Shternin:2018dcn,
    author = "Shternin, P. S. and Baldo, M. and Haensel, P.",
    title = "{In-medium enhancement of the modified Urca neutrino reaction rates}",
    eprint = "1807.06569",
    archivePrefix = "arXiv",
    primaryClass = "astro-ph.HE",
    doi = "10.1016/j.physletb.2018.09.035",
    journal = "Phys. Lett. B",
    volume = "786",
    pages = "28--34",
    year = "2018"
}

@article{vanDalen:2003zw,
    author = "van Dalen, E. N. E. and Dieperink, A. E. L. and Tjon, J. A.",
    title = "{Neutrino emission in neutron stars}",
    eprint = "nucl-th/0303037",
    archivePrefix = "arXiv",
    reportNumber = "KVI-1599, JLAB-THY-03-51",
    doi = "10.1103/PhysRevC.67.065807",
    journal = "Phys. Rev. C",
    volume = "67",
    pages = "065807",
    year = "2003"
}

@article{Bacca:2008yr,
    author = "Bacca, S. and Hally, K. and Pethick, C. J. and Schwenk, A.",
    title = "{Chiral effective field theory calculations of neutrino processes in dense matter}",
    eprint = "0812.0102",
    archivePrefix = "arXiv",
    primaryClass = "nucl-th",
    reportNumber = "NORDITA-2008-57",
    doi = "10.1103/PhysRevC.80.032802",
    journal = "Phys. Rev. C",
    volume = "80",
    pages = "032802",
    year = "2009"
}

@article{DehghanNiri:2016cqm,
    author = "Dehghan Niri, A. and Moshfegh, H. R. and Haensel, P.",
    title = "{Nuclear correlations and neutrino emissivity from the neutron branch of the modified Urca process}",
    doi = "10.1103/PhysRevC.93.045806",
    journal = "Phys. Rev. C",
    volume = "93",
    number = "4",
    pages = "045806",
    year = "2016"
}

@article{Blaschke:1995va,
    author = "Blaschke, D. and Ropke, G. and Schulz, H. and Sedrakian, A. D. and Voskresensky, D. N.",
    title = "{Nuclear in-medium effects and neutrino emissivity of neutron stars}",
    journal = "Mon. Not. Roy. Astron. Soc.",
    volume = "273",
    pages = "596--602",
    year = "1995"
}

@article{Ericson:1988wr,
    author = "Ericson, Torleif Erik Oskar and Mathiot, J. F.",
    title = "{Axion Emission from SN 1987a: Nuclear Physics Constraints}",
    reportNumber = "CERN-TH-5227/88, IPNO/TH-88-63",
    doi = "10.1016/0370-2693(89)91103-9",
    journal = "Phys. Lett. B",
    volume = "219",
    pages = "507--514",
    year = "1989"
}

@article{Mignani:2012mm,
    author = "Mignani, R. P. and Putte, D. Vande and Cropper, M. and Turolla, R. and Zane, S. and Pellizza, L. J. and Bignone, L. A. and Sartore, N. and Treves, A.",
    title = "{The birthplace and age of the isolated neutron star RX J1856.5-3754}",
    eprint = "1212.3141",
    archivePrefix = "arXiv",
    primaryClass = "astro-ph.HE",
    doi = "10.1093/mnras/sts627",
    journal = "Mon. Not. Roy. Astron. Soc.",
    volume = "429",
    pages = "3517",
    year = "2013"
}

@article{Motch:2009nq,
    author = "Motch, C. and Pires, A. M. and Haberl, F. and Schwope, A. and Zavlin, V. E.",
    title = "{Proper motions of thermally emitting isolated neutron stars measured with Chandra}",
    eprint = "0901.1006",
    archivePrefix = "arXiv",
    primaryClass = "astro-ph.HE",
    doi = "10.1051/0004-6361/200811052",
    journal = "Astron. Astrophys.",
    volume = "497",
    pages = "423",
    year = "2009"
}

@article{Tetzlaff:2011kh,
    author = "Tetzlaff, Nina and Eisenbeiss, Thomas and Neuhaeuser, Ralph and Hohle, Markus Matthias",
    title = "{The origin of RXJ1856.5-3754 and RXJ0720.4-3125 -- updated using new parallax measurements}",
    eprint = "1107.1673",
    archivePrefix = "arXiv",
    primaryClass = "astro-ph.GA",
    doi = "10.1111/j.1365-2966.2011.19302.x",
    journal = "Mon. Not. Roy. Astron. Soc.",
    volume = "417",
    pages = "617",
    year = "2011"
}

@article{Tetzlaff:2012rz,
    author = "Tetzlaff, Nina and Schmidt, Janos G. and Hohle, Markus M. and Neuhaeuser, Ralph",
    title = "{Neutron stars from young nearby associations the origin of RXJ1605.3+3249}",
    eprint = "1202.1388",
    archivePrefix = "arXiv",
    primaryClass = "astro-ph.GA",
    doi = "10.1071/AS11057",
    journal = "Publ. Astron. Soc. Austral.",
    volume = "29",
    pages = "98",
    year = "2012"
}

@article{Thorsett:2003xy,
    author = "Thorsett, S. E. and Benjamin, R. A. and Brisken, Walter F. and Golden, A. and Goss, W. Miller",
    title = "{Pulsar psr b0656+14, the monogem ring, and the origin of the `knee' in the primary cosmic ray spectrum}",
    eprint = "astro-ph/0306462",
    archivePrefix = "arXiv",
    doi = "10.1086/377682",
    journal = "Astrophys. J. Lett.",
    volume = "592",
    pages = "L71--L74",
    year = "2003"
}

@article{Zharikov:2021llh,
    author = "Zharikov, S. and Zyuzin, D. and Shibanov, Yu. and Kirichenko, A. and Mennickent, R. E. and Geier, S. and Cabrera-Lavers, A.",
    title = "{PSR B0656+14: the unified outlook from the infrared to X-rays}",
    eprint = "2101.07459",
    archivePrefix = "arXiv",
    primaryClass = "astro-ph.HE",
    doi = "10.1093/mnras/stab157",
    journal = "Mon. Not. Roy. Astron. Soc.",
    volume = "502",
    number = "2",
    pages = "2005--2022",
    year = "2021"
}

@software{2016ascl.soft09009P,
       author = {{Page}, Dany},
        title = "{NSCool: Neutron star cooling code}",
 howpublished = {Astrophysics Source Code Library, record ascl:1609.009},
         year = 2016,
        month = sep,
          eid = {ascl:1609.009},
       adsurl = {https://ui.adsabs.harvard.edu/abs/2016ascl.soft09009P}
}

@article{Hardy:2024gwy,
    author = "Hardy, Edward and Sokolov, Anton and Stubbs, Henry",
    title = "{Supernova bounds on new scalars from resonant and soft emission}",
    eprint = "2410.17347",
    archivePrefix = "arXiv",
    primaryClass = "hep-ph",
    doi = "10.1007/JHEP04(2025)013",
    journal = "JHEP",
    volume = "04",
    pages = "013",
    year = "2025"
}

@article{Friman:1979ecl,
    author = "Friman, B. L. and Maxwell, O. V.",
    title = "{Neutron Star Neutrino Emissivities}",
    doi = "10.1086/157313",
    journal = "Astrophys. J.",
    volume = "232",
    pages = "541--557",
    year = "1979"
}

@article{Bottaro:2024ugp,
    author = "Bottaro, Salvatore and Caputo, Andrea and Fiorillo, Damiano F. G.",
    title = "{Neutrino emission in cold neutron stars: Bremsstrahlung and modified urca rates reexamined}",
    eprint = "2406.18640",
    archivePrefix = "arXiv",
    primaryClass = "hep-ph",
    reportNumber = "CERN-TH-2024-092",
    doi = "10.1088/1475-7516/2024/11/015",
    journal = "JCAP",
    volume = "11",
    pages = "015",
    year = "2024"
}

@article{Page:2010aw,
    author = "Page, Dany and Prakash, Madappa and Lattimer, James M. and Steiner, Andrew W.",
    title = "{Rapid Cooling of the Neutron Star in Cassiopeia A Triggered by Neutron Superfluidity in Dense Matter}",
    eprint = "1011.6142",
    archivePrefix = "arXiv",
    primaryClass = "astro-ph.HE",
    doi = "10.1103/PhysRevLett.106.081101",
    journal = "Phys. Rev. Lett.",
    volume = "106",
    pages = "081101",
    year = "2011"
}

@article{Hamaguchi:2018oqw,
    author = "Hamaguchi, Koichi and Nagata, Natsumi and Yanagi, Keisuke and Zheng, Jiaming",
    title = "{Limit on the Axion Decay Constant from the Cooling Neutron Star in Cassiopeia A}",
    eprint = "1806.07151",
    archivePrefix = "arXiv",
    primaryClass = "hep-ph",
    reportNumber = "UT-18-13, IPMU 18-0111, IPMU-18-0111",
    doi = "10.1103/PhysRevD.98.103015",
    journal = "Phys. Rev. D",
    volume = "98",
    number = "10",
    pages = "103015",
    year = "2018"
}

@article{Sedrakian:2015krq,
    author = "Sedrakian, Armen",
    title = "{Axion cooling of neutron stars}",
    eprint = "1512.07828",
    archivePrefix = "arXiv",
    primaryClass = "astro-ph.HE",
    doi = "10.1103/PhysRevD.93.065044",
    journal = "Phys. Rev. D",
    volume = "93",
    number = "6",
    pages = "065044",
    year = "2016"
}

@article{Leinson:2014ioa,
    author = "Leinson, L. B.",
    title = "{Axion mass limit from observations of the neutron star in Cassiopeia A}",
    eprint = "1405.6873",
    archivePrefix = "arXiv",
    primaryClass = "hep-ph",
    doi = "10.1088/1475-7516/2014/08/031",
    journal = "JCAP",
    volume = "08",
    pages = "031",
    year = "2014"
}

@article{Gomez-Banon:2024oux,
    author = "G\'omez-Ba\~n\'on, Antonio and Bartnick, Kai and Springmann, Konstantin and Pons, Jos\'e A.",
    title = "{Constraining Light QCD Axions with Isolated Neutron Star Cooling}",
    eprint = "2408.07740",
    archivePrefix = "arXiv",
    primaryClass = "hep-ph",
    doi = "10.1103/PhysRevLett.133.251002",
    journal = "Phys. Rev. Lett.",
    volume = "133",
    number = "25",
    pages = "251002",
    year = "2024"
}

@article{Cowan:2010js,
    author = "Cowan, Glen and Cranmer, Kyle and Gross, Eilam and Vitells, Ofer",
    title = "{Asymptotic formulae for likelihood-based tests of new physics}",
    eprint = "1007.1727",
    archivePrefix = "arXiv",
    primaryClass = "physics.data-an",
    doi = "10.1140/epjc/s10052-011-1554-0",
    journal = "Eur. Phys. J. C",
    volume = "71",
    pages = "1554",
    year = "2011",
    note = "[Erratum: Eur.Phys.J.C 73, 2501 (2013)]"
}

@article{Dessert:2019dos,
    author = "Dessert, Christopher and Foster, Joshua W. and Safdi, Benjamin R.",
    title = "{Hard X-ray Excess from the Magnificent Seven Neutron Stars}",
    eprint = "1910.02956",
    archivePrefix = "arXiv",
    primaryClass = "astro-ph.HE",
    reportNumber = "LCTP-19-25",
    doi = "10.3847/1538-4357/abb4ea",
    journal = "Astrophys. J.",
    volume = "904",
    number = "1",
    pages = "42",
    year = "2020"
}

@article{Antypas:2022asj,
    author = "Antypas, D. and others",
    title = "{New Horizons: Scalar and Vector Ultralight Dark Matter}",
    eprint = "2203.14915",
    archivePrefix = "arXiv",
    primaryClass = "hep-ex",
    reportNumber = "FERMILAB-PUB-22-262-AD-PPD-T",
    month = "3",
    year = "2022"
}

@article{Carenza:2021pcm,
    author = "Carenza, Pierluca and Lucente, Giuseppe",
    title = "{Supernova bound on axionlike particles coupled with electrons}",
    eprint = "2107.12393",
    archivePrefix = "arXiv",
    primaryClass = "hep-ph",
    doi = "10.1103/PhysRevD.104.103007",
    journal = "Phys. Rev. D",
    volume = "104",
    number = "10",
    pages = "103007",
    year = "2021",
    note = "[Erratum: Phys.Rev.D 110, 049901 (2024)]"
}

@article{Fiorillo:2025sln,
    author = "Fiorillo, Damiano F. G. and Pitik, Tetyana and Vitagliano, Edoardo",
    title = "{Supernova production of axion-like particles coupling to electrons, reloaded}",
    eprint = "2503.15630",
    archivePrefix = "arXiv",
    primaryClass = "hep-ph",
    month = "3",
    year = "2025"
}

@article{Ho:2006uk,
    author = "Ho, Wynn C. G. and Kaplan, David L. and Chang, Philip and van Adelsberg, Matthew and Potekhin, Alexander Y.",
    title = "{Magnetic Hydrogen Atmosphere Models and the Neutron Star RX J1856.5-3754}",
    eprint = "astro-ph/0612145",
    archivePrefix = "arXiv",
    reportNumber = "SLAC-PUB-12255",
    doi = "10.1111/j.1365-2966.2006.11376.x",
    journal = "Mon. Not. Roy. Astron. Soc.",
    volume = "375",
    pages = "821--830",
    year = "2007"
}

@article{Sartore:2012fk,
    author = "Sartore, N. and Tiengo, A. and Mereghetti, S. and De Luca, A. and Turolla, R. and Haberl, F.",
    title = "{Spectral monitoring of RX J1856.5-3754 with XMM-Newton. Analysis of EPIC-pn data}",
    eprint = "1202.2121",
    archivePrefix = "arXiv",
    primaryClass = "astro-ph.HE",
    doi = "10.1051/0004-6361/201118489",
    journal = "Astron. Astrophys.",
    volume = "541",
    pages = "A66",
    year = "2012"
}

@article{Hambaryan:2011bu,
    author = "Hambaryan, V. and Suleimanov, V. and Schwope, A. D. and Neuhaeuser, R. and Werner, K. and Potekhin, A. Y.",
    editor = {G\"og\"us, Ersin and Ertan, \"Unal and Belloni, Tomaso},
    title = "{Phase resolved spectroscopic study of the isolated neutron star RBS 1223 (1RXS J130848.6+212708)}",
    eprint = "1108.3897",
    archivePrefix = "arXiv",
    primaryClass = "astro-ph.SR",
    doi = "10.1063/1.3629512",
    journal = "AIP Conf. Proc.",
    volume = "1379",
    number = "1",
    pages = "195--196",
    year = "2011"
}

@article{Hambaryan:2017wvm,
    author = {Hambaryan, V. and Suleimanov, V. and Haberl, F. and Schwope, A. D. and Neuh\"auser, R. and Hohle, M. and Werner, K.},
    title = "{The compactness of the isolated neutron star RX J0720.4\ensuremath{-}3125}",
    eprint = "1702.07635",
    archivePrefix = "arXiv",
    primaryClass = "astro-ph.HE",
    doi = "10.1051/0004-6361/201630368",
    journal = "Astron. Astrophys.",
    volume = "601",
    pages = "A108",
    year = "2017"
}

@article{Pires:2019qsk,
    author = "Pires, A. M. and Schwope, A. D. and Haberl, F. and Zavlin, V. E. and Motch, C. and Zane, S.",
    title = "{A deep XMM-Newton look on the thermally emitting isolated neutron star RX J1605.3+3249}",
    eprint = "1901.08533",
    archivePrefix = "arXiv",
    primaryClass = "astro-ph.HE",
    doi = "10.1051/0004-6361/201834801",
    journal = "Astron. Astrophys.",
    volume = "623",
    pages = "A73",
    year = "2019"
}

@article{Wilks:1938dza,
    author = "Wilks, S. S.",
    title = "{The Large-Sample Distribution of the Likelihood Ratio for Testing Composite Hypotheses}",
    doi = "10.1214/aoms/1177732360",
    journal = "Annals Math. Statist.",
    volume = "9",
    number = "1",
    pages = "60--62",
    year = "1938"
}

@article{Liu:2025ows,
    author = "Liu, Hongkai and Ohayon, Ben and Shtaif, Omer and Soreq, Yotam",
    title = "{Probing new hadronic forces with heavy exotic atoms}",
    eprint = "2502.03537",
    archivePrefix = "arXiv",
    primaryClass = "hep-ph",
    month = "2",
    year = "2025"
}

@article{Grossman:2025cov,
    author = "Grossman, Yuval and Yu, Bingrong and Zhou, Siyu",
    title = "{Axion forces in axion backgrounds}",
    eprint = "2504.00104",
    archivePrefix = "arXiv",
    primaryClass = "hep-ph",
    month = "3",
    year = "2025"
}

@article{Dev:2020eam,
    author = "Dev, P. S. Bhupal and Mohapatra, Rabindra N. and Zhang, Yongchao",
    title = "{Revisiting supernova constraints on a light CP-even scalar}",
    eprint = "2005.00490",
    archivePrefix = "arXiv",
    primaryClass = "hep-ph",
    doi = "10.1088/1475-7516/2020/08/003",
    journal = "JCAP",
    volume = "08",
    pages = "003",
    year = "2020",
    note = "[Erratum: JCAP 11, E01 (2020)]"
}

@article{Gonzalez:2010ta,
    author = "Gonzalez, Denis and Reisenegger, Andreas",
    title = "{Internal Heating of Old Neutron Stars: Contrasting Different Mechanisms}",
    eprint = "1005.5699",
    archivePrefix = "arXiv",
    primaryClass = "astro-ph.HE",
    reportNumber = "000-001",
    doi = "10.1051/0004-6361/201015084",
    journal = "Astron. Astrophys.",
    volume = "522",
    pages = "A16",
    year = "2010"
}

@article{Akmal:1998cf,
    author = "Akmal, A. and Pandharipande, V. R. and Ravenhall, D. G.",
    title = "{The Equation of state of nucleon matter and neutron star structure}",
    eprint = "nucl-th/9804027",
    archivePrefix = "arXiv",
    doi = "10.1103/PhysRevC.58.1804",
    journal = "Phys. Rev. C",
    volume = "58",
    pages = "1804--1828",
    year = "1998"
}
